\documentclass[reprint,superscriptaddress,preprintnumbers,amsmath,amssymb,aps,prd,tightenlines,longbibliography,nofootinbib]{revtex4-2}

\usepackage{graphicx}
\usepackage{xcolor}
\usepackage[sort&compress]{natbib}
\usepackage{amsmath,amssymb,bm,bbm,slashed,subdepth,amsfonts}
\usepackage{xr-hyper}
\usepackage[colorlinks=true
,urlcolor=blue
,anchorcolor=blue
,citecolor=blue
,filecolor=blue
,linkcolor=red
,menucolor=blue
,linktocpage=true
,pdfproducer=medialab
,pdfa=true
]{hyperref}
\usepackage{cleveref}
\usepackage{enumerate}
\usepackage{epsfig, subfigure}
\usepackage{setspace}
\usepackage{booktabs, tabularx}
\usepackage{units}
\usepackage{placeins}
\usepackage{multirow}
\usepackage{mathtools}
\usepackage[normalem]{ulem}
\usepackage{braket}

\usepackage{tikz}
\usetikzlibrary{shapes,arrows,positioning,automata,backgrounds,calc,er,patterns,decorations.markings,decorations.pathreplacing,snakes,decorations.pathmorphing}

\def\centerarc[#1](#2)(#3:#4:#5){ \draw[#1] ($(#2)+({#5*cos(#3)},{#5*sin(#3)})$) arc (#3:#4:#5); } 

\tikzset{
photon/.style={decorate, decoration={snake}},
gluon/.style={decorate, draw=black,
    decoration={coil,amplitude=4pt, segment length=5pt}}
 }

\newcommand{\be}{\begin{equation}}
\newcommand{\ee}{\end{equation}}
\newcommand{\bea}{\begin{equation}\begin{aligned}}
\newcommand{\eea}{\end{aligned}\end{equation}}
\newcommand{\mb}[1]{{\mathbf #1}}
\newcommand{\mc}[1]{{\mathcal #1}}

\begin{document}

\preprint{CERN-TH-2023-155}

\title{Graviton detection and the quantization of gravity}
 

\author{Daniel Carney}
\email{carney@lbl.gov}
\affiliation{Physics Division, Lawrence Berkeley National Laboratory, 1 Cyclotron Road, Berkeley, CA 94720, USA}

\author{Valerie Domcke}
\email{valerie.domcke@cern.ch}
\affiliation{Theoretical Physics Department, CERN, 1 Esplanade des Particules, CH-1211 Geneva 23, Switzerland}

\author{Nicholas L. Rodd}
\email{nick.rodd@cern.ch}
\affiliation{Theoretical Physics Department, CERN, 1 Esplanade des Particules, CH-1211 Geneva 23, Switzerland}

\begin{abstract}
We revisit a question asked by Dyson: ``Is a graviton detectable?'' We demonstrate that in both Dyson's original sense and in a more modern measurement-theoretic sense, it is possible to construct a detector sensitive to single gravitons, and in fact a variety of existing and near-term gravitational wave detectors can achieve this. However, while such a signal would be consistent with the quantization of the gravitational field, we draw on results from quantum optics to show how the same signal could just as well be explained via classical gravitational waves. We outline the kind of measurements that would be needed to demonstrate quantization of gravitational radiation and explain why these are substantially more difficult than simply counting graviton clicks or observing gravitational noise in an interferometer, and likely impossible to perform in practice.
\end{abstract}

\maketitle

A wide array of detector architectures which are sensitive to single photons now exist~\cite{2009NaPho...3..696H,PhysRevX.8.021003}. On the contrary, designing a detector capable of detecting single gravitons may seem like an insurmountable challenge. Dyson, and separately Rothman and Boughn, have argued that this would be fundamentally impossible in certain architectures, but possible with others in a sense we make precise below~\cite{Dyson:2013hbl,Rothman:2006fp,Boughn:2006st}. Supposing that one could construct such a detector, an important question arises: would observation of a signal attributable to single gravitons imply that the gravitational field is quantized? 

First, let us motivate the idea that single graviton detection is in principle possible, following Dyson. The energy density in a gravitational wave is $\rho = \tfrac{1}{4} M_{\rm Pl}^2 \langle \partial_t h^{\mu \nu} \partial_t h_{\mu \nu} \rangle$, with $M_{\rm Pl}$ the reduced Planck mass~\cite{Maggiore:2007ulw}.\footnote{We take $\hbar = c = 1$ and signature $(-,+,+,+)$ throughout.} For a wave of strain $h$ and frequency $\omega$, this corresponds to $\rho = \tfrac{1}{4} h^2 \omega^2 M_{\rm Pl}^2$. Dividing this energy up into gravitons of energy $\omega$, we find that the number of quanta per de Broglie volume $\lambda_{\rm dB}^3 = f^{-3}$ is
\be
n \lambda_{\rm dB}^3 = \frac{\pi h^2 M_{\rm Pl}^2}{2f^2} \simeq  2 \times 10^{35} \left( \frac{h}{10^{-22}} \right)^2 \left( \frac{1~{\rm kHz}}{f} \right)^{2}\!,
\label{eq:ngrav}
\ee
with $f = \omega/2\pi$ is the linear frequency. These parameters are benchmarked to a LIGO event, where it is clear that the number density of gravitons in an observable wave is astronomically large. However, it is possible to look for gravitons at much higher frequencies, where this occupation density is diluted. 

\begin{figure}[!t]
\centering
\includegraphics[width=0.45\textwidth]{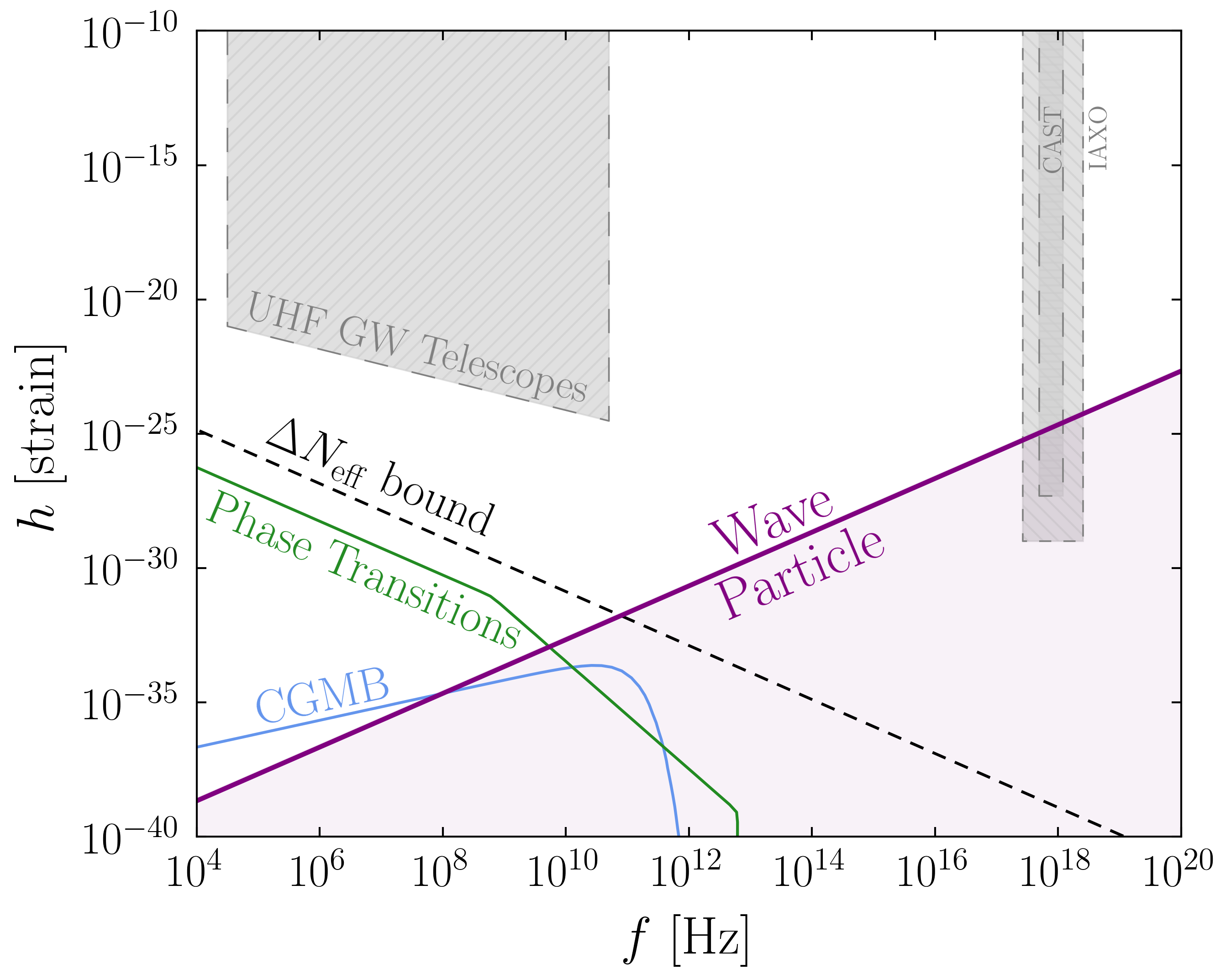}
\vspace{-0.4cm}
\caption{The parameter space for gravitational radiation at high frequency, and the region where the field -- if quantized -- would be better described as a highly dilute state of graviton particles rather than a gravitational wave [$n \lambda_{\rm dB}^3 < 1$ from Eq.~\eqref{eq:ngrav}]. Instruments such as CAST are already reaching into this parameter space.}
\label{fig:wave-pcle}
\vspace{-0.4cm}
\end{figure}

For example, consider the CERN Axion Solar Telescope (CAST)~\cite{CAST:2017uph}. This device consists of a large permanent magnet $B \sim 10~{\rm T}$ with an X-ray photodetector. The coupling $\mc{L} = \tfrac{1}{2} h_{\mu\nu} T^{\mu\nu}_{\rm EM}$ leads to linear conversion between gravitational and electromagnetic radiation in the presence of the magnetic field~\cite{Gertsenshtein,Raffelt:1987im,Ejlli:2019bqj,Berlin:2021txa,Domcke:2022rgu}. Given incident gravitational radiation, a detected X-ray photon implies the absorption of an individual graviton, under the assumption that the gravitational field is quantized. Moreover, at X-ray frequencies $f \sim 10^{18}~{\rm Hz}$ and with CAST's nominal sensitivity $h \sim 10^{-27}$~\cite{Ejlli:2019bqj}, this detector is sensitive to signals in the regime where the number quanta per de Broglie cell can become less than one. In such a state, the distribution of gravitons is highly dilute,\footnote{The analogue of Eq.~\eqref{eq:ngrav} for dark matter is $n \lambda_{\rm dB}^3 = (2\pi)^3 \rho/m^4 v^3$, so that locally dark matter undergoes the analogous transition at $m_{\scriptscriptstyle \textrm{DM}} \sim 10~{\rm eV}$. For a review, see Ref.~\cite{Hui:2021tkt}.} and in this sense CAST can detect individual gravitons.\footnote{We know of no sufficiently strong sources for gravitational radiation at this frequency (see App.~\ref{app:CAST-details}). In this paper, we will only be concerned with questions of principle involving detector sensitivities.} 

In Fig.~\ref{fig:wave-pcle} we show that the region where $n \lambda_{\rm dB}^3 < 1$ covers a large portion of the parameter space for high frequency gravitational radiation, and that CAST and its successor IAXO~\cite{Armengaud:2014gea} clearly cut into that space. The remaining aspects of the figure largely follow Ref.~\cite{Aggarwal:2020olq}, and we refer there for details, see also Refs.~\cite{Ringwald:2020ist,Goryachev:2021zzn,Herman:2020wao,Berlin:2021txa,Berlin:2023grv,Ito:2022rxn,Domcke:2022rgu,Schmieden:2022ewq,Tobar:2022pie,Domcke:2023bat,Bringmann:2023gba,Bao:2019ghe} for recent work. We emphasize that for transient signals, the strain must be interpreted carefully as discussed in App.~\ref{app:CAST-details}; in particular, the figure assumes a signal duration comparable to or longer than the measurement time, which for CAST is roughly a year.

While an intensity detector like CAST could be sensitive to such single graviton signals, it has also been suggested that linear gravitational wave detectors, for example LIGO, could be sensitive to  analogous ``shot noise'' signals due to gravitational quantization in highly squeezed states~\cite{Guerreiro:2019vbq,Parikh:2020kfh,Parikh:2020fhy,Guerreiro:2021qgk,Guerreiro:2023gdy}. Supposing that such a signal was observed in either case, a natural question to ask is whether it would constitute experimental proof that the gravitational field is quantized.\footnote{Assuming the events are not attributed to a background event or an axion. For example, in CAST, a significant background source is cosmic-ray muons interacting with the detector material and generating fluorescent X-rays ~\cite{CAST:2017uph,Armengaud:2014gea}.} The answer is no: classical gravitational radiation can equally well explain the observed detector response. 

To understand why, let us review a classic argument from quantum optics, following Ref.~\cite{mandel1995optical}. Consider a photodetector operating based on the photoelectric effect, such as a collection of approximately free electrons in a semiconductor. A fully quantized Hamiltonian describing a given detector electron coupled to the incident electromagnetic field is
\be
\label{eq:h-q}
\hat{H}_{\rm Q} = \hat{H}_{\rm det} + \hat{H}_{A} - \frac{e}{m} \hat{\mb{p}} \cdot \hat{\mb{A}},
\ee
where the first term describes a ground state and conduction band separated by an energy gap $\Delta > 0$, $\hat{\mb{p}}$ is the momentum operator of an electron, $\hat{\mb{A}}$ is the gauge field operator, and $\hat{H}_A$ is its kinetic energy. The probability for a single electron to be excited from the ground state into the conduction band in a short time $\delta t$ computed from Eq.~\eqref{eq:h-q} is,
\be
\label{eq:photoelectric}
P(t) \delta t \simeq \eta \braket{I(t)} \Theta(\omega - \Delta) \delta t.
\ee
Here, $\eta$ is a constant representing the detector efficiency, and we assume the incident field is nearly monochromatic with frequency $\omega$ and a slowly varying intensity $I(t)$. The result exhibits the hallmarks of the photoelectric effect: zero emission for $\omega < \Delta$, and emission at a rate proportional to the light intensity beyond this. Moreover, the individual events can be resolved as discrete ``clicks'', occuring when an individual electron is excited.

In contrast, consider a semi-classical model, where the gauge field is \emph{not} quantized, but viewed as a classical field:
\be
\label{eq:h-sc}
\hat{H}_{\rm SC} = \hat{H}_{\rm det} + H_A - \frac{e}{m} \hat{\mb{p}} \cdot \mb{A}.
\ee
Here $\mb{A}$ is a time dependent c-number that in principle can be stochastic, drawn from a classical distribution $P_{\rm cl}(\mb{A})$. The detector electrons remain quantized, so our experiment will continue to register a series of discrete clicks. In fact, a straightforward calculation in perturbation theory (again see Ref.~\cite{mandel1995optical} or App.~\ref{app:sc-pe}) yields precisely the same result as Eq.~\eqref{eq:photoelectric} for the rate of electron excitations. In this latter case, the bracket in $\braket{I(t)}$ is an average over the classical distribution $P_{\rm cl}(\mb{A})$.\footnote{The presence of the energy conservation $\Theta(\omega-\Delta)$ may seem surprising. In the semi-classical calculation, this arises from a factor $\left| \int_0^{\delta t} dt\,e^{i(E+\Delta-\omega)t} \right|^2 \sim \delta(E+\Delta-\omega)$ in the perturbative transition amplitude for a given electron to be excited into a conduction state of energy $E$ by an incident plane wave, $\mb{A} \propto e^{-i\omega t}$. For $\omega \delta t \gg 1$, certainly the case for example in optical detection with $\omega \gtrsim 10^{15}~{\rm Hz}$, the approximation as a delta function is excellent. More details are provided in App.~\ref{app:sc-pe}.} We conclude that the photoelectric effect can be equally well explained by a classical electromagnetic field or quantized photons.

This leads to a final question: what kind of data stream could we observe in the detector output that can be explained by a quantum model of the radiation like Eq.~\eqref{eq:h-q} but \emph{not} by a classical model like Eq.~\eqref{eq:h-sc}? A famous example in quantum optics is sub-Poisson counting statistics, in which the variance in the observed photon or graviton count rate is smaller than the mean rate \cite{mandel1995optical,davidovich1996sub}. Heuristically, if the radiation were quantized, one could send a stream of photons with fixed separation between them, leading to zero variance amongst the arrival times at the detector, and for an ideal detector, zero variance in the measured response. As we review in Sec. \ref{sec:measurement-main}, sub-Poisson counting statistics can be explained by a quantized radiation field prepared in a variety of non-trivial quantum states (e.g., a squeezed state or number eigenstate), but cannot be explained by a classical field. Thus, observation of such a signal would constitute proof that the incident field is quantized.

However, as we will emphasize, the requirements for observing such a quantization signature for gravitational radiation are very stringent. We first need some source of non-classical graviton states, and then we need a detector which can see the subtle quantum effects that differentiate this state from a classical ensemble. In particular, simply counting gravitons, or simply observing gravitational shot noise (including that produced by a highly squeezed state \cite{Guerreiro:2019vbq,Parikh:2020kfh,Parikh:2020fhy,Guerreiro:2021qgk,Guerreiro:2023gdy}) is not enough to demonstrate quantization. And even if a source for such states exists in nature, it is \emph{substantially} more difficult to make a detector capable of teasing out the subtle quantum parts of these signatures, and probably impossible in practice.

The rest of the paper is dedicated to making the above statements precise. We begin in Sec.~\ref{sec:measurement-main} by sketching a simple model of a gravitational radiation detector for illustrative purposes. We then use this to discuss the various quantization witnesses described above. In Sec.~\ref{sec:examples} we then discuss what kinds of experimental parameters would be required to construct a detector sensitive to the quantization measures. Finally in Sec.~\ref{sec:final} we relate this to other tests of quantization of gravity, for example involving the microwave background or tabletop entanglement experiments.

\section{Graviton measurement theory}
\label{sec:measurement-main}

The gravitational analogue to our models of quantized and semiclassical radiation coupled to a detector is immediate. We assume that spacetime is nearly flat $g_{\mu\nu} \simeq \eta_{\mu\nu}$ near the detector and consider small perturbations of the total metric $g_{\mu\nu} = \eta_{\mu\nu} + h_{\mu\nu}$. We will then consider the difference between a quantized gravitational field
\be
\label{eq:h-q-grav}
\hat{H}_{\rm Q} = \hat{H}_{\rm det} + \hat{H}_{h} - \frac{1}{2} \int d^3\mb{x} \, \hat{h}_{\mu\nu} \hat{T}^{\mu\nu},
\ee
where $\hat{T}^{\mu\nu}$ is some part of the matter stress tensor operator in the detector (e.g., the mirrors in LIGO or the electromagnetic field in the CAST tube), and the classical version
\be
\label{eq:h-sc-grav}
\hat{H}_{\rm SC} = \hat{H}_{\rm det} + H_{h} - \frac{1}{2} \int d^3\mb{x}  \, h_{\mu\nu} \hat{T}^{\mu\nu},
\ee
where again $h_{\mu\nu}$ is a classically random c-number field. The quantized model in Eq.~\eqref{eq:h-q-grav} should be, as usual, interpreted as an effective field theory, valid at the low energy densities considered in this paper.

The goal is to determine how these hypotheses could be distinguished. We begin in Sec. \ref{sec:classicality} with a review of the theory of coherence and classicality conditions in a radiation field. In Sec.~\ref{sec:toy-detector} we give a toy model for a gravitational wave detector which exhibits the difference between an intensity (number) detector and a linear (amplitude) detector. These two detectors offer different observables which can be used to distinguish Eqs.~\eqref{eq:h-q-grav} and \eqref{eq:h-sc-grav}, and we study these in Sec.~\ref{sec:intensity} and \ref{sec:amplitude}, respectively.

\subsection{Classicality vs. quantization}
\label{sec:classicality}

The radiation fields can be decomposed as usual into modes, whether we are dealing with classical or quantized radiation. We will find it simplest to discretize the spectrum by placing everything in a large box of length $L$, and take $L \to \infty$ at the end. In terms of positive and negative frequency components,
\bea
\label{eq:modes-main}
A_{\mu}(x) & = \sum_{\mb{k},s} \epsilon_{s,\mu}(\mb{k}) u_{\mb{k}}(\mb{x}) e^{i \omega_{\mb{k}} t} a_{\mb{k},s}^\dag + {\rm h.c.}, \\
h_{\mu\nu}(x) & = M_{\rm Pl}^{-1} \sum_{\mb{k},s} \epsilon_{s,\mu\nu}(\mb{k}) v_{\mb{k}}(\mb{x}) e^{i \omega_{\mb{k}} t} b_{\mb{k},s}^\dag + {\rm h.c.},
\eea
where the $\epsilon_s$ with $s=1,2$ form bases for the polarization vectors and tensors, and the $u_{\mb{k}},\,v_{\mb{k}}$ form complete sets of mode functions. In this section, everything will be treated symmetrically between the electromagnetic and gravitational case. Since we are primarily interested in gravity, we use the gravitational $b_{\mb{k},s}$ for the Fourier components, but the same results all apply for electromagnetism $b \to a$. See App.~\ref{app:detector} for more details on these mode expansions.

The distinction between classical and quantum models comes from how we treat the coefficients $b_{\mb{k},s}$. In the classical case, the $b_{\mb{k},s}$ are c-number dynamical variables. A general state consists of a probability distribution $P_{\rm cl}(\{ b_{\mb{k},s}\})$, which can for example include classical correlations between the different modes. In the quantum case, the $b_{\mb{k},s}$ are promoted to operators. A general state consists of a density matrix $\rho$, which can include both classical correlations and quantum correlations (entanglement). Our goal is to understand what kind of observations can distinguish between these two cases.

For simplicity, we will focus on narrow-band signals in what follows; we discuss generalizations in the appendices. Consider a single mode of the radiation field of interest (gravitational or electromagnetic), with wave vector $\mb{k}$, polarization $s$, and let $b = b_{\mb{k},s}$.\footnote{In the case of a classical gravitational wave, $b$ translates to the strain $h$ as follows: the wave has fixed energy density $\rho = \omega^2 M_{\rm Pl}^2 h^2/4 = \omega \sum_s|b_s|^2/4L^3$, so for a single polarization $b_s^2 = \omega M_{\rm Pl}^2 h_s^2 L^3$.}  We can give a simple description of the general quantum state of this mode which will make it easy to compare to the classical case. Define the coherent states $\ket{\beta}$ for this mode as usual, $b \ket{\beta} = \beta \ket{\beta}$, for any complex number $\beta$. These form an overcomplete basis for the Hilbert space of the mode, since they are non-orthogonal $\braket{\beta | \beta'} = e^{-|\beta-\beta'|^2/2}$. In fact, one can express any density matrix of the mode as
\be
\label{eq:glauber}
\rho_{\rm rad} = \int d\beta\, P(\beta) \ket{\beta} \bra{\beta}\!, \ \ \int d \beta\,P(\beta) = 1.
\ee
This representation of the density matrix is called the Glauber-Sudarshan (or sometimes ``Glauber P'') representation~\cite{Sudarshan:1963ts,Glauber:1963tx}. The normalization condition on $P(\beta)$ simply enforces ${\rm Tr}[\rho_{\rm rad}]=1$.

The representation Eq.~\eqref{eq:glauber} is highly suggestive: it looks like we can write any state of the mode as a classical ensemble of coherent states. Indeed, in the case that we \emph{can} write the radiation state as such an ensemble, one can always reproduce a detector's output with the purely classical radiation model, namely an ensemble of classical plane waves with distribution $P(\beta)$ \cite{Glauber:1963tx,mollow1975pure,mandel1995optical}. This is true \emph{even including} observation of ``vacuum fluctuations'' or shot noise, which can be explained by the quantum mechanical nature of the detector even with a purely classical signal, as discussed in the introduction. 

However, in a general quantum state, the weight function $P(\beta)$ can take negative values, or be a highly singular function.\footnote{By a theorem of Schwartz, these singular $P$-functions must take negative values. For further details, see App.~\ref{app:nonclassical-example}.} In this case, $P(\beta)$ is not a classical probability distribution. It is precisely in this case -- when $P(\beta)$ fails to be a classical distribution -- that one can find observables in a detector output that can be explained by a quantized field model like Eqs.~\eqref{eq:h-q} or \eqref{eq:h-q-grav}, but cannot explained by a classical field model like in Eqs.~\eqref{eq:h-sc} or \eqref{eq:h-sc-grav}. We therefore refer to such states as ``non-classical.'' Examples of such states include number eigenstates, superpositions of coherent states such as $\ket{\psi} \sim \ket{\beta_1} + \ket{\beta_2}$, and squeezed coherent states; see App.~\ref{app:nonclassical-example}. Notice in particular that the vacuum is classical in this definition, since it is the trivial ensemble consisting of just the zero-amplitude coherent state.\footnote{As originally shown by Glauber~\cite{Glauber:1963tx}, $P(\beta)$ obeys an analogue of the central limit theorem, implying that Gaussian $P(\beta)$, which are positive definite, are highly generic. For example, thermal radiation can be described by a Gaussian $P(\beta)$.} 

An illuminating example of such a non-classicality observable can be constructed in the optics setting, where the radiation field can be controlled. Consider an incoming beam with annihilation operator $b_1$ prepared in a coherent state $\ket{\beta}_1$. After passing this beam through a 50-50 beam splitter with vacuum $\ket{0}_2$ sent through the other port, the state of the two arms is
\be
\ket{\beta} \ket{0} \to \ket{\beta/\sqrt{2}} \ket{\beta/\sqrt{2}}\!.
\label{eq:coherent-split}
\ee
For completeness, we provide the calculation in App.~\ref{app:beamsplitter}. If we have a pair of photodetectors, one for each outgoing beam, they will continue to each register half of the intensity, even for arbitrarily small beam energy $|\beta|^2 \to 0$. On the other hand, consider something ``non-classical'' like a beam of single-photon states, again with vacuum through the dark port:
\be
\ket{1} \ket{0} \to \frac{1}{\sqrt{2}} \left[ \ket{1} \ket{0} + \ket{0} \ket{1} \right]\!.
\label{eq:number-split}
\ee
Now the photodetectors will be perfectly anti-correlated, i.e., either one detector registers the photon or the other does, at each detection event. There is no classical random ensemble that can reproduce this result \cite{Glauber:1963tx}. Note that the non-classical state produced an entangled two-body state, while the coherent state did not; this is a general feature of non-classical states \cite{kim2002entanglement}. While this example uses a beamsplitter and is thus not easy to generalize to the gravitational setting, we will give analogous examples of these kinds of measurements with realistic gravitational radiation detectors in the next sections.

To summarize, in order to perform a measurement on a gravitational radiation mode that would require a non-classical description and thus demonstrate quantization of the radiation field, we have two basic requirements:
\begin{enumerate}
\item Nature has to provide a non-classical state of the radiation mode; and
\item We have to detect some signature of this non-classicality, for example, correlation statistics as just described.
\end{enumerate}
These are very stringent conditions. In this paper, we make no comments on the possible sources of such states, although see \cite{Guerreiro:2019vbq,Parikh:2020kfh,Parikh:2020fhy,Guerreiro:2021qgk,Guerreiro:2023gdy} for some ideas. Our focus will be on the second condition, which as we shall see in what follows, entails extremely challenging detector parameters.

\subsection{A simple detector model}
\label{sec:toy-detector}

In order to be precise about detecting gravitons and non-classicality observables, we will need to discuss two important issues. The first is that different detectors measure the field in different bases. As concrete examples, CAST is an intensity detector (it clicks when a graviton is absorbed), while lower-frequency axion haloscopes based on cavities, for example ADMX \cite{Crisosto:2019fcj} or HAYSTAC \cite{HAYSTAC:2018rwy}, could be operated as amplitude detectors (they measure the amplitude of a cavity mode). Similarly, LIGO is an amplitude detector. The second issue is that, in general, real detectors do not perform projective measurements but more general, ``weak'' measurements on the incoming radiation field. This is particularly important in the gravitational case, where the incident field is extremely weakly coupled to the detector.

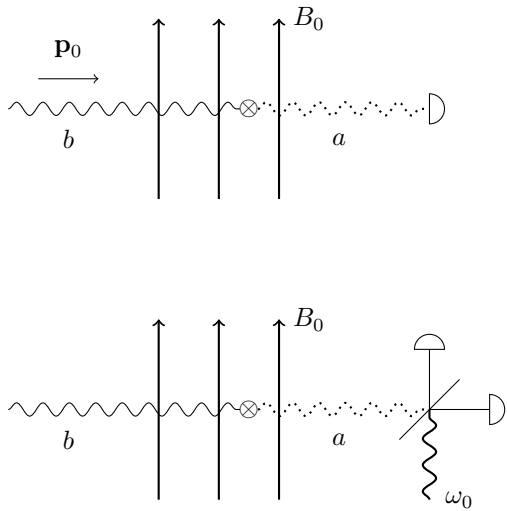
\begin{figure}[t]


\begin{tikzpicture}[scale=0.4]


\draw[dotted,photon,thick] (1.3,0) -- (7,0);
\node at (4,-1) {$a$};

\draw [thick,->] (-2,-3) -- (-2,3);
\draw [thick,->] (0,-3) -- (0,3);
\draw [thick,->] (2,-3) -- (2,3);

\node at (3,3) {$B_0$};

\draw [black] (7,0.5) -- (7,-0.5) arc(-90:90:.5) -- cycle;

\draw[photon] (-7,0) -- (0.7,0);
\node at (1,0) {$\otimes$};
\node at (-5,-1) {$b$};

\draw[->] (-6,1) -- (-4,1);
\node at (-5,2) {$\mb{p}_0$};


\begin{scope}[shift={(0,-10)}]

\draw[dotted,photon,thick] (1.3,0) -- (7,0);
\node at (4,-1) {$a$};

\draw [thick,->] (-2,-3) -- (-2,3);
\draw [thick,->] (0,-3) -- (0,3);
\draw [thick,->] (2,-3) -- (2,3);

\node at (3,3) {$B_0$};

\draw[photon] (-7,0) -- (0.7,0);
\node at (1,0) {$\otimes$};
\node at (-5,-1) {$b$};


\draw (6,-1) -- (8,1);

\draw [black] (6.5,2) -- (7.5,2) arc(0:180:.5) -- cycle;
\draw (7,0) -- (7,2);

\draw [black] (9,0.5) -- (9,-0.5) arc(-90:90:.5) -- cycle;
\draw (7,0) -- (9,0);

\draw[photon,thick] (7,-3) -- (7,0);
\node at (8,-3) {$\omega_0$};

\end{scope}

\end{tikzpicture}

\caption{Schematic implementation of our toy model detectors. The gravitational mode $b$ is incident on a fixed magnetic field $B_0$. In this field, $b$ can convert to an electromagnetic mode $a$, which can then be detected. The caps at the end represent photon counters. Top: directly counting the converted photons produces a gravitational intensity detector, as in Eq.~\eqref{eq:povm-number}. Bottom: an external light beam at frequency $\omega_0$ and known phase is interfered with the light coming out of the detector field. This produces a homodyne (amplitude) measurement of the light amplitude $X$, and thus an amplitude measurement of the gravitational signal, as in Eq.~\eqref{eq:povm-amp}.}
\label{fig:detectors}
\end{figure}

To make our discussion of both of these issues concrete, consider a simple gravitational radiation detector consisting of a strong homogeneous magnetic field $B_0$ along the $z$-axis, distributed over some spatial volume $\ell^3$, as shown in Fig.~\ref{fig:detectors}. For conceptual simplicity, we will consider an incoming ``beam'' of gravitational radiation along the $x$-axis with momentum $\mb{k}_0 = k_0 \hat{\mb{x}}$. The modes of the electromagnetic field couple to the gravitational field through the Gertsenshtein effect (i.e., the coupling $\mc{L} = \tfrac{1}{2} h_{\mu\nu} T^{\mu\nu}_{\rm EM}$) as discussed in the introduction.\footnote{Note that this interaction is quite general. In addition to its clear connection to devices like CAST, it can also describe the basic detection mechanism of interferometers like LIGO. While these are commonly described in terms of gravitational waves moving test mirrors, they can equivalently be described in terms of gravitational waves acting on the laser between the mirrors \cite{melissinos2010response}.} The interaction Hamiltonian, in the Schr\"{o}dinger picture, is
\bea
V = \,&B_0 \int_{\ell^3} \hspace{-0.12cm} d^3\mb{x}\, \left( F_{yx} h_{yy} + F_{zx} h_{yz} \right) \\
= \,&\frac{B_0}{2} \int_{\ell^3} \hspace{-0.12cm} d^3\mb{x}\, \left(\delta B_z h_+ - \delta B_y h_{\times} \right) e^{i k_0 x}+\textrm{h.c.},
\eea
where $\delta B_{y,z}$ denote the perturbations of the magnetic field strength around the background field, and $h_s = b_s/M_{\rm Pl} \sqrt{\omega L^3}$ are the amplitudes of the individual polarizations, according to the conventions we establish in App.~\ref{app:detector}. We include this here to emphasize that each polarization couples to a specific direction of the magnetic field perturbations.

Suppose that we can monitor a particular mode $\mb{p}_0,r_0$ of the electromagnetic field, with annihilation operator $a = a_{\mb{p}_0,r_0}$ and frequency $\omega_0 = |\mb{p}_0|$, polarized with $B$ along the $y$-axis for definiteness. Again for simplicity, we will assume that this mode is perfectly matched to the incoming gravitational radiation $\mb{p}_0 = \mb{k}_0$. In this case, the interaction Hamiltonian can be simplified to
\be
V = i g \big(a^\dagger b - a b^\dagger\big), \ \ \ g = \frac{B_0}{2 \sqrt{2} M_{\rm Pl}} \sqrt{\frac{\ell^3}{L^3}}.
\label{eq:V-main}
\ee
The coupling $g$ has units of energy. The artificial spatial scale $L$ reflects the discretized radiation modes $b$; in the continuum limit these scale like $L^{3/2}$ and one would essentially replace $L \to \delta k^{-1}$ in $g$ to get the effective coupling to a radiation signal of bandwidth $\delta k$. A more general and rigorous form of this discussion is provided in App.~\ref{app:detector}. Our simplifying assumptions here make this the ``best case'' scenario: we have a narrowband, perfectly mode-matched, optimally orientated detector. We will see that even in this case, detection of non-classicality witnesses is extraordinarily difficult; in a more realistic case the difficulties will only increase.

First we consider how to use this as an intensity detector. We assume that at each time step of length $\delta t$, the electromagnetic mode $a$ is prepared in its vacuum state $\ket{0}$, and we then measure the mode $a$ in its number basis $\ket{n}$, for example with a photodiode. Suppose to begin with that the gravitational radiation mode is in some pure state $\ket{\psi_{\rm rad}}$, so that the initial total state is $\ket{0,\,\psi_{\rm rad}}$. The probability that we see $n$ photons excited in the detector mode in the timestep $\delta t$ is given by
\bea
\label{eq:pn}
p(n) & = \sum_{f_{\rm rad}} | \braket{n,\, f_{\rm rad}| U | 0,\, \psi_{\rm rad}} |^2 \\
& = \braket{ \psi_{\rm rad} | K_n^\dag K_n | \psi_{\rm rad}}\!,
\eea
where the sum over the final state of the gravitational field $\ket{f_{\rm rad}}$ arises because we do not measure it directly. Here
\be
\label{eq:UVI}
U = \exp \left[ -i \int_0^{\delta t} dt\, V_I(t) \right]
\ee
is the interaction-picture evolution operator, which encodes the interaction between the radiation and detector. We have used this to define the Krauss operator 
\be
K_n = \braket{ n | U | 0}\!,
\ee
which acts on the radiation field. These Krauss operators generalize projective measurements, which can be recovered when the operators $E_n = K_n^\dag K_n$ are projectors.\footnote{In measurement theory, the operators $E_n = K_n^\dag K_n$ form what is called a positive operator-valued measure (POVM) \cite{nielsen2010quantum}. They satisfy $\sum_n E_n = 1$ and therefore give an operator-valued measure on the space of outcomes, in this case, values of $n$ that can be registered by the detector. If the radiation is in a mixed state $\rho_{\rm rad}$, then Eq.~\eqref{eq:pn} is generalized to $p(n) = {\rm tr} [E_n \rho_{\rm rad}]$.} The gravitational coupling is so weak that we are instead making a very weak, non-projective measurement. For example, with our specific coupling Eq.~\eqref{eq:V-main}, we have
\be
\label{eq:povm-number}
K_n^\dag K_n = (1 - g^2 \delta t^2 b^{\dagger} b)\, \delta_{n,0} + g^2 \delta t^2 b^\dag b\, \delta_{n,1},
\ee
up to terms of $\mc{O}(g^4 \delta t^4)$. What this means is that with overwhelming probability, the detector will have zero excited photons, while the next most likely outcome, with probability $g^2 \delta t^2 \braket{\psi_{\rm rad} | b^\dag b | \psi_{\rm rad}}$, is that a single detector photon is excited. 

Essentially, this says that our detector is a graviton version of a photodiode: it is excited with probability proportional to the incoming intensity $I \sim \omega b^\dag b$. Explicitly, in the model where $b$ is a classical c-number, we find that the probability of a single detector excitation is
\be
\label{eq:p-1}
p(1) = g^2 \delta t^2 |b|^2 = \frac{1}{2} \frac{\rho}{\omega} \frac{B_0^2 \ell^3}{M_{\rm Pl}^2} \delta t^2,
\ee
using Eq.~\eqref{eq:V-main}. Thus, the conversion probability is proportional to the number density of incident gravitons, $\rho/\omega$. This expression is reminiscent of the Gertsenshtein effect, and indeed we will show the connection is exact.

The same basic detector architecture can also be used to implement linear (``amplitude,'' ``quadrature'') measurements on the gravitational radiation mode. We define the electromagnetic field quadrature operators
\be
X = \frac{a + a^\dag}{\sqrt{2}},\hspace{0.5cm}Y = -i \frac{a - a^\dag}{\sqrt{2}},
\ee
which are canonically conjugate $[X,Y] = i$ and act like position and momentum operators; these are sometimes called the amplitude and phase variables. Suppose that instead of monitoring the electromagnetic mode $a$ with a photocounting device, we monitor it with something like a homodyne interferometer (see Fig.~\ref{fig:detectors}). This means that we projectively measure it in one of the (continuous) quadrature bases, say the amplitude basis $\ket{X}$. Again we assume that we do this repeatedly every time step $\delta t$.\footnote{\label{phase-note}We are making another important simplifying assumption here, which is that we know the phase of the incoming signal. In a more realistic setting, say with an astrophysical source, the unknown relative phase $\phi$ between the signal and the detector mode would have to appear either in the coupling $g \to g e^{i \phi}$ in Eq.~\eqref{eq:V-main} or in the measurement basis $X \to X(\phi) = (a e^{i \phi} + a^\dag e^{-i \phi})/\sqrt{2}$. As with our other simplifying assumptions, this assumption will only make the task of detecting field quantization \emph{easier}, and since as we will see this is already impossibly difficult, we will continue to assume a phase-matched detector for a simpler presentation.} The probability distribution for the detector to be measured with value $X$ is\footnote{Note that in this equation $\ket{0}$ is still the vacuum $n=0$ state, not the amplitude $X=0$ eigenstate.}
\be
\label{eq:pofX}
p(X) = \braket{\psi_{\rm rad} | K_X^\dag K_X | \psi_{\rm rad}}\!, \ \ \ K_X = \braket{X | U | 0}.
\ee
Again using our specific interaction Hamiltonian Eq.~\eqref{eq:V-main}, we find to linear order in $g \delta t$,
\be
\label{eq:povm-amp}
K_X^\dag K_X  = f_0(X) + g \delta t f_1(X) [ b + b^\dag],
\ee
where
\bea
f_0(X) & = \left| \braket{X|0} \right|^2, \\
f_1(X) & = \braket{0 | X} \braket{X | 1}\!,
\eea
involve the position-space harmonic oscillator wavefunctions $\braket{X | n}$, and we used the fact that $f_1(X)$ is real. We see that the detector can in principle be excited to any value of $X$. The probability distribution on these detector $X$ outcomes is shifted by an amount proportional to the \emph{gravitational radiation field} quadrature $X_{\rm b} = (b + b^\dag)/\sqrt{2}$. For example, with an incident coherent state $\ket{\psi_{\rm rad}} = \ket{\beta}$, the average value on the detector is $\overline{X} = \sqrt{2} g \delta t {\rm Re}(\beta) \propto h$, where $h$ is the usual strain observable. Thus this constitutes a gravitational amplitude detector.

We emphasize that a gravitational radiation intensity detector is one that measures in the number basis. In this sense one can ``detect single gravitons'' simply by building such a detector. As we will explain next, traditional quantum optics experiments demonstrating quantization of the photon have used this kind of detector, operated in a regime where the incident light is intense (i.e., $n \lambda_{\rm dB}^3 \gg 1$). Therefore, to detect a quantization measure, it is \emph{not} necessary to be detecting in the dilute beam limit $n \lambda_{\rm dB}^3 \ll 1$, contrary to the implicit suggestion in Dyson's original work~\cite{Dyson:2013hbl}. Conversely, detecting a signal in this dilute limit does not imply or require a quantized description of the signal field.

\subsection{Intensity measurements}
\label{sec:intensity}

We now turn to intensity measurements on the gravitational radiation and how they can be used to test the quantization of the gravitational field. What we will do is calculate the most general signal possible with the classical model in Eq.~\eqref{eq:h-sc-grav}, then perform the analogous computation with the quantized model of Eq.~\eqref{eq:h-q-grav}, and then show some examples of intensity data that can be explained in the quantized but not the classical model. Following our discussion above, we will model the detector data as a series of clicks: in each interval $\delta t$, either $0$ or $1$ click occurs. Over a total integration time $T \gg \delta t$, we can then ask for the probability of observing $N$ clicks, $p(N)$.

In the classical case, Eqs.~\eqref{eq:h-sc} or \eqref{eq:h-sc-grav}, the general state of the radiation mode is described by a probability distribution $P_{\rm cl}(b)$ for the Fourier amplitude. First consider an incoming state with a fixed value of $b$. Assuming that the coupling $g$ in Eq.~\eqref{eq:V-main} is weak $g \delta t \ll 1$, the only measurement outcomes in each timestep $\delta t$ with reasonable probability are zero clicks (with $p(0) = 1 - g^2 \delta t^2 |b|^2$), and one click, with $p(1)= g^2 \delta t^2 |b|^2$ from Eq.~\eqref{eq:p-1}.
The average number of clicks in time $T$ is then
\be
\label{eq:PoissonMean}
\overline{N} = p(1) \frac{T}{\delta t} = \eta \phi T,
\ee
where we have defined
\be
\label{eq:phieta}
\phi = \frac{\rho \ell^2}{\omega} = \frac{|b|^2 \ell^2}{4 L^3}, \ \ \ \eta = 4 g^2 \frac{L^3}{\ell^3} \ell \delta t = \frac{B_0^2}{2 M_{\rm Pl}^2} \ell \delta t,
\ee
which represent the incoming flux $\phi$ and detector efficiency $\eta$. (In the quantum picture, $\phi$ corresponds to the number of gravitons incident on the detector per unit time, whereas the efficiency determines the fraction of those gravitons that are absorbed.) 

Within a long time interval $T \gg \delta t$, we imagine collecting $N$ total clicks. Because we are rapidly resetting the detector at each time step, the individual detection events are independent, and so will be Poisson distributed, and thereby fully described by the mean in Eq.~\eqref{eq:PoissonMean}. Thus we have
\be
p(N|\,b) = \frac{1}{N!} [\eta \phi T]^{N} e^{-\eta \phi T}.
\ee
The variance in the number of clicks is then given by the usual Poisson formula,
\be
\label{eq:var-deterministic}
\Delta N^2 = \overline{N}.
\ee
Notice, in particular, that the signal $b$ here is completely deterministic; this variance represents randomness in the detector purely due to the detector's quantum fluctuations.

In the most general classical model, $b$ is not fixed, but instead is a random variable drawn from a distribution $P_{\rm cl}(b)$. As a result, the flux is stochastic, and therefore so too is the expected number of events in a small time interval, $\eta \phi(b) \delta t$, where we have made explicit the fact that the flux $\phi = \phi(b)$ depends on $b$ through Eq.~\eqref{eq:phieta}. Nevertheless, the probability for observing $N$ clicks in a time $T$ can be derived just as in the Poisson case,
\bea
p(N) = &\frac{1}{N!} \int db\, P_{\rm cl}(b) [\eta \phi T]^{N} e^{-\eta \phi T}.
\eea
The click variance is now given by
\bea
\label{eq:dn2-c}
\Delta N^2 &= \overline{N} + \eta^2 T^2 \int db\, P_{\rm cl}(b) \left[ \phi - \braket{\phi} \right]^2 \\
&= \overline{N} + \eta^2 T^2 \big\langle \left[ \phi - \braket{\phi} \right]^2 \big\rangle,
\eea
where here $\braket{\cdot}$ denotes an integral over the classical distribution, so that $\braket{\phi} = \int db\, P_{\rm cl}(b) \phi(b)$. In the case where $b$ is fixed, $P_{\rm cl}(b)=\delta(b-b')$, the final term on the right hand side vanishes, reproducing Eq.~\eqref{eq:var-deterministic}. More generally, as a classical probability distribution satisfies $P_{\rm cl}(b) \geq 0$, Eq.~\eqref{eq:dn2-c} implies that $\Delta N^2 \geq \overline{N}$: the click statistics must appear either as Poisson or ``super-Poisson.'' Note, however, that as $\overline{N} \propto \eta$, whereas the super-Poisson contribution scales as $\eta^2$: the statistics revert to Poisson for $\eta \ll 1$.

Now consider the model in Eqs.~\eqref{eq:h-q} or \eqref{eq:h-q-grav}, where the incident field is quantized into photons or gravitons. Contrary to what we concluded for the classical field model, there are quantum states that can produce sub-Poisson statistics, $\Delta N^2 < \overline{N}$. We will analyze this in direct analogy with the classical case. First, consider an incoming radiation mode in a definite coherent state $\ket{\beta}$. The probability of a single click in the detector is
\be
p(1) = \braket{\beta | K_1^\dag K_1 | \beta} = g^2 \delta t^2 | \beta|^2.
\ee
Moreover, the state of the incident radiation mode after this measurement is still $\ket{\beta}$.\footnote{After a measurement described by a Krauss operator $K_a$, the state $\ket{\psi} \to K_a \ket{\psi}/ \| K_a \ket{\psi}\|$ where $a$ is the measurement outcome, which is the generalization of the usual projection postulate. In this particular case, since $K_1 \propto b$, the measurement leaves a coherent state $\ket{\beta}$ unchanged to leading order in $g \delta t \ll 1$. This is why the coherent state basis is used in the theory of photodetection.} Thus in a time interval $T \gg \delta t$, we can have $N$ clicks which will again be exactly Poisson distributed,
\be
p(N|\,\beta) = \frac{1}{N!} [\eta \phi T]^N e^{-\eta \phi T},
\ee
where now $\phi = |\beta|^2 \ell^2/4L^3$, so that $\overline{N} = \eta \phi T$. Here, $\hat{\phi} \sim b^\dag b$ is now an operator, and $\phi = \phi(\beta) = \braket{\beta | \hat{\phi} | \beta}$ is its expectation value in a specific coherent state $\ket{\beta}$. As claimed, $\phi$ is a direct measure of the number of incident gravitons per unit time. In the general case, the state of the mode is described by a density matrix $\rho_{\rm rad}$, which can be described with the Glauber-Sudarshan representation of Eq.~\eqref{eq:glauber}. In analogy to the classical case, one then finds that the variance in the observed detector counts is
\be
\label{eq:dn2-q}
\Delta N^2 = \overline{N} + \eta^2 T^2 \int d \beta\, P(\beta) \left[ \phi - \braket{\phi} \right]^2\!,
\ee
where now $\braket{\phi} = {\rm Tr}[\rho_{\rm rad} \phi]$. See App.~\ref{app:variance} for the detailed calculation. In spite of the superficial similarity to Eq.~\eqref{eq:dn2-c}, there is a key difference: in a general radiation field state, $P(\beta)$ can be negative for certain values of $\beta$, and the integral in Eq.~\eqref{eq:dn2-q} is not necessarily positive. Therefore, observation of sub-Poisson detector clicks, $\Delta N^2 < \overline{N}$, can be explained by the quantum model of the radiation field but not the classical version.

Detecting sub-Poisson fluctuations is a formidable task. This was only accomplished in the 1980's in optics \cite{short1983observation,rempe1990observation}. As discussed in Sec. \ref{sec:classicality}, there are two reasons. The first is that we need a source state that can produce a Glauber distribution which has $P(\beta) < 0$ for at least some values of $\beta$. A coherent state, where $P(\beta) = \delta(\beta-\beta')$, does not satisfy this condition, although it can be achieved for example with a phase-squeezed state (see App.~\ref{app:nonclassical-example}). However, in the gravitational case, perhaps the more important issue is that one needs a high-efficiency detector. This is clear from Eq.~\eqref{eq:dn2-q}: if $\eta \ll 1$, the actual data will be Poisson distributed. Exactly as for the classical model in Eq.~\eqref{eq:dn2-c}, any deviation from Poisson statistics is suppressed by $\eta$. Heuristically, this is a statement that the detected sample statistics are a very poor approximation to the underlying population statistics in the signal, and are dominated by the intrinsic Poisson statistics of the detector itself.

Modern photocounting devices can achieve $\eta \sim 0.9$ \cite{2009NaPho...3..696H,PhysRevX.8.021003}. What about a gravitational intensity detector? With our toy model, we can make a crude estimate of the sensitivity of a real detector like CAST. Taking $B_0 \simeq 10~{\rm T}$ and $\ell \simeq \delta t = 10~{\rm m}$, the efficiency is
\be
\label{eq:eta-numeric}
\eta = \frac{B_0^2}{2M_{\rm Pl}^2} \ell \delta t \simeq 8.2 \times 10^{-34}.
\ee
Thus, the data will always look Poissonian. In other words, there is no way for CAST to make a counting statistics measurement that can differentiate between a classical gravitational wave and a quantized beam of gravitons. While this is very suggestive that any realistic detector cannot make such a measurement, a natural question is whether it is possible in principle. We analyze this in Sec.~\ref{sec:examples}.

Finally, it is important to note that there are observables in intensity detectors other than sub-Poisson statistics that can distinguish between the classical and quantized models. Examples include antibunching as observed in $g^2(\tau)$ and higher-order autocorrelation measurements \cite{Glauber:1963tx}. One can also consider entangled states (e.g., simultaneous detection of entangled graviton pairs). We have not systematically studied all of these, but in each example, the basic issue of detector efficiency should be highly prohibitive independent of the specific observable considered.

\subsection{Amplitude measurements}
\label{sec:amplitude}

Linear detectors of radiation fields also admit tests of quantization analogous to sub-Poisson counting statistics. Consider the amplitude detector described by Eq.~\eqref{eq:povm-amp}. We read out the detector by measuring its amplitude quadrature $X$, which corresponds to a measurement of the gravitational field amplitude $X_b = (b + b^\dag)/\sqrt{2}$. As a first guess, we could try to look at the variance of this observable, in direct analogy with the counting statistics of the previous section. 

One finds a result which is very similar in spirit to the number counting statistics. The detector data will have variance given by
\be
\label{eq:varX}
\Delta X^2 = \frac{1}{2} + g^2 \delta t^2 
\begin{cases}  \braket{ \Delta X_b^2}\!, & {\rm classical}, \\
\big[ \!\braket{ \Delta X_b^2}\! - \frac{1}{2} \big], & {\rm quantum}.
\end{cases}
\ee
Writing this more explicitly, in the classical radiation case, we have
\be
\Delta X^2 = \frac{1}{2} + 2 g^2 \delta t^2 \int db\, P_{\rm cl}(b) \left[ {\rm Re}~b - \braket{ {\rm Re}~b}\right]^2\!,
\ee
while in the model with quantized radiation, we have instead
\be
\label{eq:varX-sig}
\Delta X^2 = \frac{1}{2} + 2 g^2 \delta t^2 \int d\beta\, P(\beta) \left[ {\rm Re}~\beta - \braket{ {\rm Re}~\beta}\right]^2\!.
\ee
A detailed calculation is provided in App.~\ref{app:variance}. In both the classical and quantum cases, the first $1/2$ term represents the vacuum fluctuations of the detector mode itself (i.e., noise at the ``standard quantum limit'').\footnote{As is well known, this is not a fundamental limit. The $1/2$ can in principle be reduced arbitrarily close to $0$ by using various quantum measurement techniques, such as the preparing the detector mode in a squeezed state instead of the vacuum. However, this only helps in the case that the phase of the incoming signal is known (see footnote \ref{phase-note}). Moreover, in a realistic device, squeezing can at most achieve a reduction of this factor $1/2$ by a few orders of magnitude before becoming limited by optical losses.} The second terms are essentially the variance of the gravitational mode $X_b$, which is transduced onto the detector mode $X$ with efficiency $g \delta t \ll 1$. 

Much like the case of number counting statistics, the key difference is that the integral over $P_{\rm cl}(b)$ is positive-definite, while the integral over the quantum $P(\beta)$ can be negative. In particular, this means that with particular incoming quantum states---for example, a highly squeezed gravitational mode---it is possible that the detector's data will actually \emph{decrease} in variance compared to its native vacuum fluctuations. This behavior is impossible with an incoming classical signal.

However, observing this effect is probably impossible in practice, because $g \delta t$ is astronomically small in a real detector. In particular, the maximum quantum effect, even with an infinitely squeezed state, would be to render $\Delta X^2 = 1/2 - g^2 \delta t^2$. In this case, the gravitational amplitude is fluctuating less than in its vacuum state. This is directly analogous to the counting statistics: quantum mechanics allows for radiation states which will produce ``less noisy'' data than any classical state (because they can act to reduce the noise in the detector itself), but observing this requires overcoming a detector efficiency $\eta \sim g \delta t \ll 1$.

In Refs.~\cite{Parikh:2020kfh,Parikh:2020fhy}, it was argued that observing shot noise induced from a highly squeezed, large-amplitude gravitational state would constitute evidence for the quantization of gravity. We agree that a highly squeezed state can lead to an exponential enhancement of the measured strain noise (see App.~\ref{app:nonclassical-example}), where $\braket{\Delta X_b^2} \gg 1$. However, observing this enhanced noise would not provide any information about the quantization of the field, since either the quantum or classical radiation models can produce the same data. This does not require an exotic or highly tuned classical model; a simple Gaussian distribution of classical plane waves with the same width $\braket{\Delta X^2_b}$ would suffice to reproduce the same signal. The only ``uniquely quantum'' signal would be if the minute fluctuations of the noise have $\braket{\Delta X_b^2} < 1/2$, i.e., if the gravitational amplitude is fluctuating less than its vacuum state. As we have argued above, observing this would be extremely challenging with a realistic detector.

For completeness, we also note that there are other quadrature observables that can distinguish between the classical and quantized cases. In principle, for example, one could fully reconstruct the underlying density matrix using a tomographic set of measurements \cite{lvovsky2009continuous,Guerreiro:2021qgk}, which in particular would allow one to check if $P(\beta) < 0$. However, we expect any of these observables to be similarly limited by the detector efficiency issues discussed here.

\section{Can the quantization signatures be detected in principle?}
\label{sec:examples}

In the previous section, we described experimental measurements which could clearly demonstrate that the gravitational radiation field is quantized. Even if we allowed for sources of the appropriate non-classical graviton states, all such measurements appear prohibitively difficult with realistic terrestrial detectors. The fundamental obstruction was encoded in Eq.~\eqref{eq:eta-numeric}: signatures of non-classicality are highly suppressed by the miniscule detection efficiency. In this section, we turn to the question of whether it is possible even in principle, given the known laws of physics, to construct a detector capable of observing these subtle quantization effects in the gravitational radiation field. 

To do so, let us return to the example of an amplitude detector based on graviton to photon conversion, i.e.\ a futuristic version CAST. Naively, the basic question is whether the laws of physics will allow us to increase $B_0$ and/or the detector size $\ell$ such that the detector efficiency $\eta \approx 1$. In this case, the explicit challenge is the small graviton-photon conversion probability, $p(g \to \gamma)$, which is controlled by the inverse Gertsenshtein effect, and well studied as a classical wave mixing problem, see Refs.~\cite{Raffelt:1987im,Ejlli:2018hke,Domcke:2020yzq}. The probability that a graviton converts to a photon over the length of the detector, however, is exactly the detector efficiency in the language of Eq.~\eqref{eq:phieta}, with $\delta t = \ell$ set by how long the graviton takes to traverse the detector. Accordingly, we can also compute the efficiency from the inverse Gertsenshtein effect,
\be
\eta = \frac{B_0^2}{2M_{\rm Pl}^2} \ell_{\rm osc}^2 \sin^2\left( \frac{\ell}{\ell_{\rm osc}}\right)\!.
\label{eq:prob_G}
\ee
Here $\ell$ denotes the length over which the oscillation can occur within a homogeneous magnetic field $B_0$ orthogonal to the propagation direction, exactly as in our discussion in Sec.~\ref{sec:measurement-main}.

However, this general result includes an additional effect: self-interactions of the electromagnetic field, as encoded in the Euler-Heisenberg Lagrangian. These imply that increasing $B_0$ will not arbitrarily enhance the detection efficiency. Instead, a larger $B_0$ eventually reduces the coherence of the photon and graviton, thereby reducing the mixing. This is captured by the oscillation length $\ell_{\rm osc}$ that is determined from $4/\ell_{\rm osc}^2 = m_{\rm EH}^4/\omega^2 + 2B_0^2/M_{\rm Pl}^2$, where for a single polarization $m_{\rm EH}^2 = 7 \omega^2 (e^2/180 \pi^2) (B_0/B_{\rm crit})^2$ and $B_{\rm crit} = m_e^2/e$ with $m_e$ and $e$ the electron mass and charge, respectively. The result as stated neglects any possible plasma mass for the photon, and further assumes $\omega \gg m_{\rm EH}$, which is justified for all parameters we will consider.

For CAST, $B_0 = 10~{\rm T}$, $f = 10^{18}~{\rm Hz}$, and $\ell=10~{\rm m}$, so that in natural Heaviside-Lorentz units 
\be
\frac{2B_0^2}{M_{\rm Pl}^2} \simeq \left(5.6~{\rm pc} \right)^{-2}\!,\hspace{0.5cm}
\frac{m_{\rm EH}^4}{\omega^2} \simeq \left( 0.17~{\rm AU} \right)^{-2}\!.
\ee
Accordingly $\ell_{\rm osc} \simeq 0.35~{\rm AU} \gg \ell$, implying that the Euler-Heisenberg corrections are irrelevant. Indeed, when $\ell \ll \ell_{\rm osc}$ the efficiency in Eq.~\eqref{eq:prob_G} reduces to Eq.~\eqref{eq:eta-numeric}, as it must, demonstrating the equivalence of the classical and quantum derivation of the Gertsenshtein process. If we seek to maximize $\eta$, $\ell_{\rm osc}$ must be accounted for. Doing so for $\ell = 1~{\rm AU}$, the maximum efficiency occurs for $B_0 \simeq 6~{\rm T}$, but remains small, $\eta \simeq 5 \times 10^{-14}$. Over cosmological distances sizeable values can be achieved: for $\ell = 1~{\rm Mpc}$ ($1~{\rm Gpc}$) and $B_0 = 0.1~{\rm G}$ ($1~{\rm mG}$) we can achieve $\eta \simeq 1\%$ ($70\%$). While such large devices are certainly unrealistic, we do not see a physical law that would prohibit such an instrument existing in the Universe, thereby allowing $\eta \sim \mc{O}(1)$. The observation that CAST has already reached the dilute graviton regime, while even a Gpc version barely meets the requirements to test quantization underlines how much more difficult the latter task is.

\section{Outlook and relation to other tests}

\label{sec:final}

Gravity can be consistently quantized at low energy densities as an effective quantum field theory \cite{Donoghue:1994dn,Donoghue:2022eay}. An important question is to determine what kinds of measurements could be used to verify that this is the way gravity really operates in nature. We have analyzed Dyson's suggestion that one could try to make a detector that is sensitive to single gravitons. The conclusion is straightforward: it is possible to make such a detector, but simply seeing the detector click due to incoming gravitational radiation could be explained either with quantized gravitons or simple classical gravitational waves. Vice versa, from the point of view of high-frequency gravitational wave searches venturing into the parameter space shown in Fig.~\ref{fig:wave-pcle}, the dilute graviton regime is a red herring: crossing this line in the parameter space does not come with any particular change in the difficulty of performing a gravitational wave measurement. 

In order to make a measurement that cleanly distinguishes between the quantum and classical cases, two very substantial hurdles have to be overcome: a source of non-classical (e.g., squeezed or entangled) gravitational radiation has to be produced, and then detected with a gravitational detector with high efficiency. We do not see a realistic path to satisfying either of these conditions, although we also have no argument that nature is fundamentally incapable of satisfying them.

There are two known alternatives to testing low-energy quantum gravity. One is to look at fluctuations in the cosmic microwave background, assuming these are produced by a phase of cosmic inflation. As the arguments in this paper should make clear, observation of a Gaussian power spectrum (in either the scalar or tensor fluctuations) can easily be explained by either classical or quantized gravitational radiation. Similarly, the measurement of CMB tensor modes with superhorizon correlations (as proposed, e.g., in Ref.~\cite{krauss2014using}), without independent evidence pointing to vacuum fluctuations of inflation, can also be explained by either hypothesis. 
It may however be possible to produce good non-classicality witnesses with higher-point observables~\cite{maldacena2016model,Green:2020whw}. 

The other is to use tabletop experiments to observe quantum effects generated by the two-body Newton interaction $\hat{V} = G_N m_1 m_2/|\hat{\mb{x}}_1 - \hat{\mb{x}}_2|$ (for a review, see Ref.~\cite{carney2019tabletop}). It is known that a coherent, entangling Newton interaction is only consistent with unitarity and Lorentz invariance if gravitons also exist in the spectrum of scattering states \cite{belenchia2018quantum,carney2022newton}. Thus experimental verification that the low-energy interaction is coherent and unitary would provide an indirect proof that the radiation field is also quantized. We view the clear difficulties with measurements directly on the gravitational radiation presented in this paper as substantial motivation to pursue both of these programs.

\section*{Acknowledgements}

We thank Markus Aspelmeyer, Diego Blas, Camilo Garcia-Cely, Daniel Green, Thiago Guerriero, Giacomo Marocco, Maulik Parikh, Grant Remmen, Jess Riedel, Mariana Safronova, John Selby, Yotam Soreq, Raman Sundrum, Jacob Taylor, Jure Zupan, and Kathryn Zurek for discussions. DC is supported in part by the U.S. Department of Energy QuantISED program, grant KA2401032, and by the Heising-Simons foundation, grant 2023-4467.

\bibliographystyle{utphys}
\bibliography{refs}

\appendix

\section{CAST as a single graviton detector}
\label{app:CAST-details}

Figure~\ref{fig:wave-pcle} is suggestive that CAST has the required sensitivity to detect individual gravitons. Here we demonstrate that this is indeed the case. We will also see that it is extremely difficult to conceive of a suitable source close to the current sensitivity of CAST. Of course, we emphasize, as demonstrated in the main body, even a detection would a priori not prove the source was quantized.

To begin with, the CAST analysis in Ref.~\cite{Ejlli:2019bqj} reports a characteristic strain sensitivity of $h_c \simeq 5 \times 10^{-28}$ to a stochastic gravitational wave background at $10^{18}\,{\rm Hz}$. This can be translated to an energy density, $\rho = \rho_c \int d\ln \omega\, \Omega_{\rm gw}$, with $\Omega_{\rm gw} = \omega^2 h_c^2 M_{\rm Pl}^2/2\rho_c$ and $\rho_c$ the critical energy density. The CAST sensitivity thus corresponds to the following energy and number density,
\bea
\rho &\simeq 2 \times 10^{18}~{\rm keV/cm}^3, \\
n &\simeq 4 \times 10^{17}~{\rm gravitons/cm}^3.
\eea
Although these densities are large, the de Broglie volume is not, and we have $n \lambda_{\rm dB}^3 \sim 10^{-5} \ll 1$. Hence, a (weak) signal in CAST due to GWs would imply GW detection in the single graviton limit.

However, a stochastic gravitational wave background at this amplitude is firmly excluded by the constraints on extra radiation from BBN and CMB~\cite{Planck:2018vyg,Pisanti:2020efz,Yeh:2020mgl}. We thus proceed to estimate the sensitivity of CAST to transient signals, which are not subject to these bounds. Let us consider a transient signal with frequency $f = 10^{18}\,{\rm Hz}$, amplitude $h$, and duration $T_{\rm gw}$. Then, comparing the energy density of a plane GW with the CAST limit yields a sensitivity of 
\be
h \simeq  7 \times 10^{-28} (T_m/T_{\rm gw})^{1/4},
\ee
where $T_m \simeq 1$~year denotes the measurement time of CAST and the factor $(T_{\rm gw}/T_m)^{1/4} \leq 1$ accounts for finite duration of the GW signal assuming a statistics limited experiment. As we are now considering a transient rather than stochastic signal we have swapped to $h = \sqrt{2} h_c$, which matches between the conventions of the present work and Ref.~\cite{Ejlli:2019bqj}. Consequently, the CAST sensitivity implies a detection threshold of $n \lambda_{\rm dB}^3 \simeq 10^{-5}  (T_m/T_{\rm gw})^{1/2}$,  which means that CAST can reach the single graviton threshold for signals lasting $T_{\rm gw} \sim 4\,{\rm ms}$ at $h \simeq 2 \times 10^{-25}$.

A source with these properties is, however, difficult to conceive. The prototypical example of an ultra high-frequency GW source is the inspiral of light primordial black holes (PBH).  Matching the CAST frequency and the required $T_{\rm gw}$ fixes $m_{\scriptscriptstyle \textrm{PBH}} \simeq 4 \times 10^{-24}~M_{\odot}$ (for equal masses in the binary). At such small masses, the PBHs would dissipate due to Hawking radiation in less than a day (and for instance are strongly excluded as being the dark matter of the Universe~\cite{Carr:2021bzv}). Even putting this aside, in order to reach the required amplitude of $h \sim 10^{-25}$ requires a distance between the binary and the detector of $\mc{O}(\textrm{cm})$, which clearly rules this out as a viable source. There are other possible sources -- for a review see Refs.~\cite{Aggarwal:2020olq,Franciolini:2022htd} -- although these appear similarly challenging at best.

In summary, while interesting that CAST does reach the single graviton regime and, in principle, one could imagine a GW with the required properties, conceiving of a viable source for such a GW is challenging. Moreover, as discussed in Sec.~\ref{sec:examples} a measurement of sub-poissonian statistics, seems completely out of reach with any realistic version of this technology: it would require an improved detection efficiency -- with the associated challenges discussed in Sec.~\ref{sec:examples} -- and the source itself having a non-classical distribution. In this sense, our conclusions are very similar to those of Dyson~\cite{Dyson:2013hbl}, though we stress that the difficulty lies in finding a viable GW source and in measuring sub-poissonian statistics, not in the single graviton detection itself.

\section{Semi-classical photoelectric effect}
\label{app:sc-pe}

Here we briefly expand on the justification that the semi-classical Hamiltonian in Eq.~\eqref{eq:h-sc} reproduces the famous observations of the photoelectric effect encoded in Eq.~\eqref{eq:photoelectric}. This is a well known result, and our discussion closely follows Ref.~\cite{mandel1995optical}. As mentioned in the main text, even when the electromagnetic field is classical, we continue to infer its presence with quantized electrons, and therefore our measurements remain discrete clicks. Further, the appearance of the intensity will remain, as the transition is dictated by quantum mechanics, it must depend on $|\mb{A}(t)|^2 \sim I(t)$. More interesting is to consider the behavior when $\omega \sim \Delta$. Working in the interaction picture, the transition probability in a small time interval $\delta t$ is given by
\bea
{\rm TP} = &\left| \int_0^{\delta t} dt\, \bra{E} \hat{H}_{{\rm SC},I}(t) \ket{-\Delta} \right|^2 \\
= &\left| \frac{e}{m}\int_0^{\delta t} dt\, \bra{E} \hat{\bf p} \ket{-\Delta} \cdot {\bf A}(t)\,e^{i(E+\Delta)t} \right|^2\!.
\eea
If we treat $\mb{A}$ as describing a nearly monochromatic plane wave with frequency $\omega$, the relevant contribution will be proportional $e^{-i \omega t}$, so that the transition probability is controlled by
\bea
\label{eq:tp-sin}
\left| \int_0^{\delta t} dt\,e^{i(E+\Delta-\omega)t} \right|^2
= \left[ \frac{\sin[\tfrac{1}{2} (E+\Delta-\omega) \delta t]}{\tfrac{1}{2}(E+\Delta-\omega)} \right]^2\!.
\eea
Even for $\omega < \Delta$, this quantity is non-zero. Nevertheless, for realistic values of $\Delta$ and $\omega$, it is highly suppressed. For $\Delta \sim \omega \sim {\rm eV} \sim 10^{15}~{\rm Hz}$, we expect $(\omega-\Delta)\delta t \gg 1$, and the final line of Eq.~\eqref{eq:tp-sin} is very well approximated by $\delta(E+\Delta-\omega)$. To determine the total probability for excitation into the conduction band, we integrate the transition probability over all $E$, finding no support when $\omega < \Delta$, consistent with Eq.~\eqref{eq:photoelectric}. The claim that the semi-classical model is fully consistent with Eq.~\eqref{eq:photoelectric} therefore follows.

\section{An example non-classical $P(\beta)$ function}
\label{app:nonclassical-example}

In the main text, we showed how a quantum state whose Glauber representation $P(\beta)$ is negative for some values of $\beta$ can produce output in a detector which cannot be reproduced by any classical radiation signal, even one with classical randomness. In this appendix, we explain how $P(\beta)$ works for specific quantum state of interest: the squeezed coherent states. We first define these states, calculate their number and amplitude variances $\braket{\Delta n^2}$ and $\braket{ \Delta X^2}$, then show how these observables can be used to deduce the negativity of $P(\beta)$.

The squeezed coherent states are Gaussian states of a harmonic oscillator whose variance in one quadrature is below the vacuum value, say $\Delta X < 1/2$, which by Heisenberg uncertainty requires that $\Delta Y > 1/2$. They can be efficiently defined in terms of the squeezing and displacement unitary operators (see Sec. 21 of Ref.~\cite{mandel1995optical}):
\bea
S(z) & = \exp \left\{ \frac{1}{2} \left[ z^* b^2 - z b^{\dag 2} \right] \right\}\!, \\
D(\beta) & = \exp \left\{ \beta b^\dag - \beta^* b \right\}\!.
\label{eq:SDdef}
\eea
Both $\beta$ and $z$ are arbitrary complex numbers. The displacement operator rotates a given coherent state $\ket{\alpha}$ into $\ket{\beta + \alpha}$. The squeezing operator takes a coherent state into a complicated superposition of coherent states which we will describe shortly. To calculate statistics in these states, it is most useful to use their transformation properties on the creation and annihilation operators:
\bea
\label{eq:transforms}
& S^\dag b S = \mu b - \nu b^\dag, \ \ S^\dag b^\dag S = \mu b^\dag - \nu^* b, \\
& D^\dag b D = b + \beta, \ \ D^\dag b^\dag D = b^\dag + \beta^*,
\eea
where the complex squeezing number $z$ is parametrized $z = r e^{i \theta}$, or alternatively
\be
\label{eq:munu}
\mu = \cosh r, \ \ \nu = e^{i \theta} \sinh r.
\ee
Often $r$ is called the squeezing amplitude and $\theta$ is the squeezing angle. In terms of these operators, the squeezed vacuum state is $\ket{z} := S(z) \ket{0}$. More generally one can consider a squeezed coherent state,\footnote{Note that $[S,D] \neq 0$. Our definition here is what has historically been called a ``two-photon squeezed state,'' while the alternative $DS\ket{0}$ states were called ``ideal squeezed states.'' This choice is somewhat arbitrary, as one can always express a two-photon squeezed state as an ideal one, or vice versa. The terminology ``two-photon'' does not mean that the states have two photons (indeed they do not have a definite number), but rather reflects the fact that to generate $S(z)$ from a Hamiltonian $S = e^{-i H t}$ requires $H$ to be bilinear in the photon operators. For further discussion see Ref.~\cite{mandel1995optical}.}
\be
\ket{z,\beta} := S(z) D(\beta) \ket{0}\!.
\ee

Generally speaking, to create a squeezed state from the vacuum or coherent state, one requires a non-linear interaction to enact the squeeze operator. These states have been produced in optical fields since the 1980's, using, for example, non-linear crystals. In the gravitational case, there is no definitively known source of squeezed states, although inflation and the non-linearities of gravitational mergers have been suggested as possible sources \cite{Guerreiro:2019vbq,Parikh:2020kfh,Parikh:2020fhy,Guerreiro:2021qgk,Guerreiro:2023gdy}.

Let us now study the nature of the ``non-classicality'' of the squeezed states. The most efficient way to understand this is by computing the observables studied in the main text, such as the signal variance $\braket{\Delta X^2}$, and then comparing the specific results to the general results we showed must hold in the classical case. Recall the definition of the amplitude quadrature $X = (b+b^\dag)/\sqrt{2}$. In a general squeezed coherent state, simple algebra using Eq.~\eqref{eq:transforms} gives the average
\be
\braket{X}_{z,\beta} = \frac{1}{\sqrt{2}} \left[ (\mu - \nu^*) \beta + (\mu - \nu) \beta^* \right]\!,
\ee
with $\braket{X}_{z,\beta} = \braket{z,\beta | X | z,\beta}$. The expectation of $X^2$ is
\be
\braket{X^2}_{z,\beta} = \frac{1}{2} |\mu - \nu|^2 \braket{b b^\dag}_0 + \braket{X}_{z,\beta}^2\!,
\ee
where we left $\braket{b b^\dag}_0 = \braket{0 | b b^\dag | 0} = 1$ explicit to highlight the ``vacuum fluctuations.'' We then obtain the variance
\be
\label{eq:x2squeezed}
\braket{\Delta X^2}_{z,\beta} = \frac{1}{2} | \mu - \nu|^2.
\ee
For example, the true vacuum is $\mu = 1, \nu = 0$ (i.e. $r = 0$) so that $\braket{\Delta X^2}_0 = 1/2$, the usual vacuum fluctuations. What squeezing does is to modify the variance. In particular, we can choose values of the squeezing parameters such that $\braket{\Delta X^2} < 1/2$ is below the vacuum value; for example a large amount of squeezing $r \to \infty$ and $\theta = 0$ sends $\braket{\Delta X^2} \to 0$. This does not violate Heisenberg uncertainty: one can perform the analogous calculation in the conjugate variable $Y = -i(b - b^\dag)/\sqrt{2}$ and find that for the same state, $\braket{\Delta Y^2} \to \infty$. In this manner, squeezing distributes all the uncertainty into a specific quadrature, whereas the standard vacuum distributes it evenly between variables. 

The result in Eq.~\eqref{eq:x2squeezed} is enough to show that the Glauber function $P(\beta)$ for such states is negative in some region. Compare this result to the general result in any Glauber state, our expression $\braket{ \Delta X_b^2 }_{\rm qu}$ in Eq.~\eqref{eq:varX-sig}. In a squeezed state where $\braket{\Delta X^2} < 1/2$, Eq.~\eqref{eq:varX-sig} immediately implies that the integral over $\beta$ must be negative, from which it follows that $P(\beta) < 0$ for some values of $\beta$. It is worth emphasizing that the non-classicality of the state is independent of the coherence $\beta$; it depends only on the squeezing parameters. 

In principle, one can look to show $P(\beta) < 0$ directly by computing the Glauber function. This turns out to be difficult. For instance, we can compute the Glauber function using a formula due to Mehta~\cite{mehta1967diagonal}, which gives $P$ in terms of the density matrix:
\be
P(\beta) = \frac{e^{|\beta|^2}}{\pi^2}  \int d\alpha\braket{-\alpha | \rho | \alpha} e^{|\alpha|^2} e^{\beta \alpha^* - \beta^* \alpha}.
\ee
For example, with a coherent state $\rho = \ket{\beta_0} \bra{\beta_0}$, this formula produces the expected $P(\beta) = \delta(\beta - \beta_0)$. However, inserting a squeezed state, one finds a highly singular distribution in which $P(\beta)$ is expressed as an infinite series of derivatives acting on Dirac delta functions $\sim (\partial/\partial \beta)^n \delta(\beta)$. A similar results holds for a Fock state $\ket{n}$ with definite graviton number. This highlights the fact that in general $P(\beta)$ is a distribution in the sense of a Dirac delta function, not an actual function. A general theorem due to Schwartz, however, shows that any distribution of order greater than zero (which roughly means that it has points which behave like derivatives of the Dirac delta) must always be negative in some region. See, for instance, Chapter 6 of Ref.~\cite{rudin1980functional}.

Finally, we mention that one can also produce sub-Poisson counting statistics with squeezed signals. Let $n = b^\dag b$ be the number operator of the signal, i.e., the graviton mode. We use $n$ to distinguish this from $N$, the number of clicks observed in the detector. Using the same algebraic tools as above, one finds the average
\bea
\braket{n}_{z,\beta} &= |\nu|^2 + |\mu \beta^* - \nu^* \beta|^2 = \sinh^2 r \\
&\hspace{-0.7cm}+|\beta|^2 \big[\cosh^2 r + \sinh^2 r - \sinh 2r \cos(2\phi-\theta)\big],
\eea
where we invoked Eq.~\eqref{eq:munu} and  wrote the coherence parameter as $\beta = |\beta| e^{i \phi}$. The variance takes more work, but one obtains
\bea
\label{eq:dn2-app1}
&\braket{\Delta n^2}_{z,\beta} = \braket{n}_{z,\beta} + \sinh^2r \cosh2r  \\
& \ \ \  + 2|\beta|^2 \sinh r \big[\sinh3r - \cosh3r \cos(\theta-2\phi) \big].
\eea
Taking $r=0$, we recover the Poissonian result, $\braket{\Delta n^2} = \braket{n}$, as expected for a coherent state. The statistics of the squeezed state are more general. To exhibit sub-Poisson variance, the final term in Eq.~\eqref{eq:dn2-app1} must dominate. Therefore, for simplicity, let us take $\theta=2\phi$, and consider a large-amplitude state $|\beta| \gg 1$. Then the above results combine as
\be
\frac{\braket{ \Delta n^2} - \braket{n}}{\braket{n}} = -1 + e^{-2r}.
\ee
By taking $r \gg 1$ (but $\ll \tfrac{1}{3} \ln|\beta|$) this can be arbitrarily close to $-1$, which would mean a state with \emph{no} fluctuations in the number flux $\braket{\Delta n^2} \to 0$, which is the most sub-Poisson distribution possible.

\section{Beamsplitters}
\label{app:beamsplitter}

A general beamsplitter is defined by a scattering matrix
\be
\begin{pmatrix} b_1' \\ b_2' \end{pmatrix} = \begin{pmatrix} t & r \\ r^* & t \end{pmatrix} \begin{pmatrix} b_1 \\ b_2 \end{pmatrix}\!,
\ee
with $b_i$ the input modes and $b_i'$ the output modes. Let $U$ be the unitary that implements $b_1 \to b_1' = U b_1 U^\dag = t b_1 + r b_2$, etc. The action on a single-excitation state is:
\bea
U \ket{10} & = U b_1^\dag U^\dag \ket{00} = (t^* b_1^\dag + r^* b_2^\dag) \ket{00} \\
& = t^* \ket{10} + r^* \ket{01}\!,
\eea
where we used $U^\dag \ket{00} = \ket{00}$.\footnote{This in turn follows because $b_1 U^\dag \ket{00} = U^\dag (U b_1 U^\dag) \ket{00} = U^\dag (t b_1 + r b_2) \ket{00} = 0$, and similarly for $b_2$.} The action on a pair of coherent states can be computed similarly. The coherent states are generated by the displacement operator of Eq.~\eqref{eq:SDdef}
\bea
\ket{\beta_1 \beta_2} & = D(\beta_1,\beta_2) \ket{00}, \\
D(\beta_1,\beta_2) & = \exp \left\{ (\beta_1b_1^\dag \!-\! \beta_1^* b_1) + (\beta_2 b_2^\dag \!-\! \beta_2^* b_2) \right\}\!. 
\eea
Writing $D$ out as a Taylor series and inserting factors of $U^\dag U = 1$, it is clear that this transforms as
\be
U D(\beta_1,\beta_2) U^\dag = D(\beta_1',\beta_2'),
\ee
where
\be
\beta_1' = t^* \beta_1 + r \beta_2, \ \  \beta_2' = t^* \beta_2 + r^* \beta_1.
\ee
Again using $U^\dag \ket{00} = \ket{00}$, we then see that an incoming pair of coherent states is transformed by the beamsplitter into a new pair of coherent states,
\be
U \ket{\beta_1 \beta_2} = U D(\beta_1,\beta_2) U^\dag \ket{00} = \ket{\beta_1',\beta_2'}\!.
\ee
Choosing $t = r = 1/\sqrt{2}$ justifies Eqs.~\eqref{eq:coherent-split} and \eqref{eq:number-split} from the main text.

\section{Detector toy model}
\label{app:detector}

In this appendix, we provide a more detailed discussion of the detector Hamiltonian of Eq.~\eqref{eq:V-main} used in the main text. As described there, the basic idea is to use is the fluctuations in a single mode of the electromagnetic field with an external, homogeneous magnetic field as the detector.

To determine the interaction between the electromagnetic detector modes and the gravitational field, we begin with the action for the coupling between electromagnetism and gravity,
\be
S = \int d^4x\, \sqrt{-g} \left( - \frac{1}{4} \, g^{\mu \alpha} g^{\nu \beta} F_{\mu \nu} F_{\alpha \beta} \right)\!. 
\ee
Expanding the metric around flat spacetime $g_{\mu\nu} = \eta_{\mu\nu} + h_{\mu\nu}$, one obtains the interaction to leading order in the fluctuations $h$:
\bea
S_{\rm int} & = \frac{1}{2} \int d^4x\, h_{\mu\nu} T^{\mu\nu}, \\
T^{\mu \nu} & = F^{\mu \alpha} F^\nu_{\hphantom{\nu} \alpha} - \tfrac{1}{4} \eta^{\mu \nu} F_{\alpha \beta} F^{\alpha \beta}, 
\eea
where here all indices are raised and lowered by $\eta_{\mu\nu}$. In the Hamiltonian framework, we can write this as an interaction Hamiltonian
\be
V = - \frac{1}{2} \int_{\rm \ell^3} \hspace{-0.12cm} d^3 \mb{x}\, h_{\mu\nu} T^{\mu\nu}.
\ee
The integration is performed over the detector volume, $\ell^3$, which is the region we assume our magnetic field is localized to.

Turning to the magnetic field, we assume it is homogeneous, has magnitude $B_0$ and is oriented along the $z$-axis. We expand $F_{\mu\nu} = F_{0,\mu\nu} + \delta F_{\mu\nu}$ into fluctuations around this background,\footnote{More precisely, we are taking the external magnetic field to be static in the transverse traceless frame, whose coordinates are given by freely falling observers. We make this assumption mainly for simplicity, although it will not impact the qualitative conclusions we reach. In general, the difference between frames for high-frequency gravitational waves must be treated carefully, see for instance the discussion in Refs.~\cite{Berlin:2021txa,Domcke:2022rgu,Domcke:2023bat}. Nevertheless, if the frequency exceeds the mechanical resonances of the system, as is the case for the setups we consider, we expect this to be a good approximation to a more complex detector model~\cite{Bringmann:2023gba,Berlin:2023grv}.} where
\be
F_{0,\mu\nu} = \begin{pmatrix} 0 & 0 & 0 & 0 \\ 0 & 0 & B_0 & 0 \\ 0 & -B_0 & 0 & 0 \\ 0 & 0 & 0 & 0 \end{pmatrix}\!.
\ee
In transverse traceless gauge, this reduces the interaction Hamiltonian to the simple form
\be
\label{eq:V-general}
\!\!V = B_0 \int_{\ell^3} \hspace{-0.12cm} d^3\mb{x}\, \Big[ h_{xz} \delta B_x + h_{yz} \delta B_y - (h_{xx} + h_{yy} ) \delta B_z \Big].
\ee
This expression is completely general and only relies on the orientation of the external magnetic field.

Now we want to determine the part of the coupling in Eq.~\eqref{eq:V-general} that drives the electromagnetic mode that we will detect. We can decompose the electromagnetic gauge field (in the Schr\"{o}dinger picture) as
\be
\label{eq:modes-A}
\delta A_{\mu}(\mb{x}) = \sum_{\mb{p},r} \epsilon_{r,\mu}(\mb{p})  u_{\mb{p}}(\mb{x}) a^\dag_{\mb{p},r} + {\rm h.c.},
\ee
where $\epsilon_r(\mb{p})$ are a complete set of polarization vectors, and the $u_{\mb{p}}(\mb{x})$ form a complete, discrete set of modes:
\be
\int_{\ell^3} \hspace{-0.12cm} d^3\mb{x}\, \ u_{\mb{p}}(\mb{x}) u_{\mb{p}'}^*(\mb{x}) = \frac{\delta_{\mb{p}, \mb{p}'}}{2 \omega_{\mb{p}}}, \ \ \ \epsilon_{r}(\mb{p}) \cdot \epsilon^*_{r'}(\mb{p}) = \delta_{r,r'}.
\ee
In general, the nature of these modes depends on the details of the detector. For a device like CAST, which detects X-ray wavelength photons ($\lambda \lesssim 1~{\rm nm}$) in a detector volume of order $m^3$, the modes of interest can be modelled as travelling plane waves. In an experiment like a microwave axion cavity, one instead detects a single mode whose wavelength is the size of the cavity, and one should really use stationary standing waves. For simplicity here, we will use plane waves, 
\be
u_{\mb{p}}(\mb{x}) = \frac{e^{-i \mb{p} \cdot \mb{x}}}{\sqrt{2  \omega_{\mb{p}} \ell^3}}, \ \ \ \mb{p} = \frac{2 \pi \mb{n} }{\ell},
\ee
with $\mb{n}$ an arbitrary vector of integers, though this result can be easily generalized to cavity modes with appropriate boundary conditions. The normalizations are chosen so that the energy density in the field modes is
\be
H_{\rm EM} = \frac{1}{2} \int_{\ell^3} \hspace{-0.12cm} d^3\mb{x} \left[ \mb{E}^2 + \mb{B}^2 \right] = \sum_{\mb{p},r} \omega_{\mb{p}} a^\dag_{\mb{p},r} a_{\mb{p},r},
\ee
plus the usual infinite zero-point energy. 

Meanwhile, the gravitational radiation field $h_{\mu\nu}$ is freely propagating no matter what the detector boundary conditions are, so we expand it as in Eq.~\eqref{eq:modes-main}, viz.,
\be
\label{eq:modes-h}
h_{\mu\nu}(\mb{x}) = M_{\rm Pl}^{-1} \sum_{\mb{k},s} \epsilon_{s,\mu\nu}(\mb{k}) v_{\mb{k}}(\mb{x}) b_{\mb{k},s}^\dag + {\rm h.c.},
\ee
where we take the polarization tensors $\epsilon_{s,\mu\nu}(\mb{k})$ to be real, and normalized according to,
\be
\epsilon_{s, \mu \nu}(\mb{k})\, \epsilon_{s'}^{\mu\nu}(\mb{k}) = \delta_{s,s'}.
\ee
Conventionally the two polarizations are labeled as $s=+,\times$.\footnote{Concretely, we work in the transverse traceless gauge, and define the polarization tensors as $\epsilon_+^{ij} = (u^i u^j - v^i v^j)/\sqrt{2}$, $\epsilon_{\times}^{ij} = (u^i v^j + v^i u^j)/\sqrt{2}$ with $\bf{v} = (\hat{\bf{e}}_z \times \hat{\bf{k}})/|\hat{\bf{e}}_z \times \hat{\bf{k}}| = \hat{\bf e}_\phi$ and $\bf{u} = \bf{v} \times \hat{\bf{k}}$.} The mode functions are just the usual plane waves, normalized again to fix the correct expression for the kinetic energy, and so that $h$ is a dimensionless strain: 
\be
v_{\mb{k}}(\mb{x}) = \frac{e^{-i \mb{k} \cdot \mb{x}}}{\sqrt{2  \omega_{\mb{k}} L^3}}, \ \ \ \mb{k} = \frac{2 \pi \mb{n} }{L}.
\ee
To connect our notation with a classical gravitational wave of a single mode, we would take amplitudes $b_s = h_s M_{\rm Pl} \sqrt{\omega L^3}$, so that 
\be
h^{\mu\nu}(x) = \frac{1}{\sqrt{2}} \big[h_+ \epsilon_+^{\mu\nu}(\mb{k}) + h_\times \epsilon_\times^{\mu\nu}(\mb{k})\big] e^{i k \cdot x} + {\rm h.c.},
\ee
and the total strain is $h^2 \equiv |h_+|^2 + |h_\times|^2$.

Now, suppose we are monitoring a single mode $a = a_{\mb{p}_0,r_0}$ of the electromagnetic field, with wavevector $\mb{p}_0 = \omega_0 \hat{\mb{x}}$ and polarization $r_0$.\footnote{If the detector is an electromagnetic cavity, this can be done easily by just isolating a single cavity frequency. For a detector with a continuum of modes, one can isolate a single mode for example with filter cavities near the final photodetectors. We use the single-mode language for simplicity, but the results generalize in a straightforward way to finite-bandwidth detectors.} We take this to be a mode such that its magnetic field at $t=0$ is aligned along the $y$-axis. Then only the $\delta B_y$ term in Eq.~\eqref{eq:V-general} contributes. Using the mode expansion of Eq.~\eqref{eq:modes-A} with $\mb{B} = \nabla \times \mb{A}$, and the gravitational mode expansion Eq.~\eqref{eq:modes-h}, we find that the interaction Hamiltonian between the mode $a$ and the gravitational field is given by
\be
\label{eq:v0}
V = \frac{B_0}{M_{\rm Pl}} \int_{\ell^3} \hspace{-0.12cm} d^3 \mb{x} \, (\mc{O}_b + \mc{O}_b^\dag) (\mc{O}_a + \mc{O}_a^\dag),
\ee
where we have defined the operators
\be
\mc{O}_b   = \sum_{\mb{k},s} \epsilon_{s,yz}(\mb{k})  v^*_{\mb{k}}(\mb{x}) b_{\mb{k},s},  \quad 
\mc{O}_a  =  i \omega_0 u^*_{\mb{p}_0}(\mb{x}) a.
\ee
Equation~\eqref{eq:v0} represents the precise coupling of the detector mode $a = a_{\mb{p}_0,r_0}$ to the entire gravitational field; notice that it involves a sum over all the gravitational modes. 

Finally, we can simplify the interaction potential by performing the volume integral. We use the integral
\bea
W(\mb{p},\mb{k}) =& \int_{-\ell/2}^{\ell/2} d^3\mb{x}\, e^{i (\mb{p}-\mb{k}) \cdot \mb{x}} \\
=& \prod_{i=x,y,z} \frac{2 \sin[(p^i - k^i) \ell/2]}{p^i-k^i}.
\eea
Again the precise nature of this window function depends on the relative size of the detector and signal mode of interest. For simplicity, consider the case where $\lambda_{\rm sig} \ll \ell$, and where we the detector integrates over a reasonable number of signal periods $\omega \delta t \gg 1$ (as in, for example, CAST). Then, $W(\mb{p},\mb{k}) \to \ell^3 \delta_{\mb{p},\mb{k}}$, and we have simply
\be
\label{vfinalapp}
V \rightarrow \frac{i B_0}{2 \sqrt{2} M_\text{Pl}} \sqrt{\frac{\ell^3}{L^3}} \big( a^\dagger b - a b^\dagger \big),
\ee
where $b = b_{\mb{p}_0,\times}$. This completes the derivation of Eq.~\eqref{eq:V-main} in the main text. 

We end with a note on our final two assumptions. We have assumed the detector is large compared to the signal wavelength; this assumption can be relaxed in a straightforward way by just leaving the window functions $W$ under an integral. Formally, the result in Eq.~\eqref{vfinalapp} also contains a contribution proportional to $a b_{-\mb{p}_0,\times}$ and its conjugate, which we have dropped by assuming $\omega \delta t \gg 1$. In the case of a detector which averages over many periods of the signal, these naturally average out in the interaction Hamiltonian Eq.~\eqref{eq:UVI} (the ``rotating wave approximation''). In a slower detector, like a resonant amplitude detector, these terms do in fact contribute by adding an additional quanta of graviton vacuum fluctuations. As we will see in App.~\ref{app:variance}, this additional quanta will not substantially change any of our conclusions, and can safely be ignored.

\section{Variance calculations}
\label{app:variance}

In this appendix we provide the detailed computations of the variances in the click rates -- Eqs.~\eqref{eq:dn2-c} and \eqref{eq:dn2-q} -- and amplitude measurements -- Eq.~\eqref{eq:varX} -- given in the main text. Our discussion partly follows Ref.~\cite{mandel1995optical}, and we refer there for further details.

\subsection{Number variance}

Consider the intensity detector discussed in Sec.~\ref{sec:intensity}. We begin with the classical calculation. First, assume a classical incoming radiation state with definite Fourier amplitude $b$. The mean number of clicks observed in a time $T$ is then fixed by Eq.~\eqref{eq:PoissonMean} to $\overline{N} = \eta \phi T$, with $\phi = \phi(b) = |b|^2 \ell^2/4L^3$ the incoming gravitational flux, as in Eq.~\eqref{eq:phieta}. Even so, the observed number of detector clicks will be stochastic. As the events will each be independent, the observed number of clicks $N$ in a single observation of time $T$ will be Poisson distributed,
\be
\label{eq:pNb}
p(N|\,b) = \frac{1}{N!} [\eta \phi(b) T]^N e^{-\eta \phi(b) T},
\ee
where we explicitly note this is the distribution in the semi-classical model. We can confirm the average is as expected,
\bea
\overline{N} & = \sum_{N=0}^{\infty} p(N|\,b) N \\
& =  \eta \phi(b) T \sum_{N=1}^{\infty} \frac{1}{(N-1)!} [\eta \phi(b) T]^{N-1} e^{-\eta \phi(b) T} \\
& = \eta \phi(b) T.
\label{eq:barN-Poisson}
\eea
Similarly,
\begin{align}
\overline{N(N-1)} & =  [\eta \phi(b) T]^2 \sum_{N=2}^{\infty} \frac{1}{(N-2)!} [\eta \phi(b) T]^{N-2} e^{-\eta \phi(b) T} \nonumber \\
& = [\eta \phi(b) T]^2,
\label{eq:barN2-Poisson}
\end{align}
from which we conclude
\bea
\Delta N^2 = \overline{N(N-1)} + \overline{N} - \overline{N}^2 = \eta \phi(b) T,
\eea
as one expects in a Poisson distribution. These are elementary calculations; we include them as the calculations in the more general scenarios will mirror these. Note that we use overlines to denote averages taken over many observations.

Now instead of a definite value for $b$, suppose that it is a classically random variable with distribution $P_{\rm cl}(b)$. Physically this could correspond to a stochasticity in the amplitude of the input wave. By the law of total probability, we now have
\bea
p(N) =& \int db\,P_{\rm cl}(b) p(N |\,b) \\
= &\frac{1}{N!} \int db\, P_{\rm cl}(b) [\eta \phi(b) T]^{N} e^{-\eta \phi(b) T}.
\label{eq:braket-pN}
\eea
Despite the superficial similarity, for a general $P_{\rm cl}(b)$ Eq.~\eqref{eq:braket-pN} is not the Poisson distribution. This can be confirmed by explicit calculation. We have
\bea
\overline{N} &= \int db\,P_{\rm cl}(b) \sum_{N=0}^{\infty} p(N |\,b) N \\
&= \int db\,P_{\rm cl}(b)\, \eta \phi(b) T \\
&= \eta T \!\braket{\phi}\!,
\eea
directly from Eq.~\eqref{eq:barN-Poisson}. Here we use the notation $\braket{\cdot}$ to denote an integral over $P_{\rm cl}(b)$ is performed, for example $\braket{\phi} = \int db P_{\rm cl}(b) \phi(b)$. We emphasize that this is distinct from expectation values over the distribution of the clicks observed in a time $T$, which we represent with an overline. Continuing, Eq.~\eqref{eq:barN2-Poisson} implies
\bea
\overline{N(N-1)} = \int db P_{\rm cl}(b) [\eta \phi(b) T]^2 = \eta^2 T^2 \!\braket{\phi^2}\!,
\eea
so that the variance is now
\bea
\Delta N^2 = \overline{N} + \eta^2 T^2 \big\langle(\phi-\braket{\phi})^2\big\rangle \geq \bar{N},
\eea
as quoted in the main text in Eq.~\eqref{eq:dn2-c}.

These same arguments can now be generalized to the case where the incident graviton flux is quantized. Consider first an incoming field of definite coherent state $\ket{\beta}$, such that the flux takes a fixed value of $\phi = \phi(\beta) = |\beta|^2 \ell^2/4L^3$. Clicks in the detector will remain independent, so that again the observed number will be Poisson distributed,
\be
p(N|\,\beta) = \frac{1}{N!} [\eta \phi(\beta) T]^N e^{-\eta \phi(\beta) T}.
\ee
The mean and variance are identical to the fixed classical case with $|b|^2 \to |\beta|^2$. 

A general quantized initial state is described by a Glauber $P(\beta)$, and although this is not a probability distribution, the general distribution is given by
\be
\label{pN-q}
p(N) = \frac{1}{N!} \int d\beta\, P(\beta)\, [\eta \phi(\beta) T]^N e^{-\eta \phi(\beta) T}.
\ee
A careful derivation justifying this form for a stationary incident field is given in Ref.~\cite{mandel1995optical}. Crucially, however, Eq. \eqref{pN-q} is \emph{not} the quantum expectation value of Eq. \eqref{eq:pNb} in the state $\ket{\beta}$, with $|b|^2 = b^\dag b$ interpreted as an operator. Rather, it is the expectation value of the \emph{normal ordered} version of $p(N | b)$. This latter statement follows from a general result sometimes called the optical equivalence theorem, which says that for an operator $\mathcal{O}(b,b^\dag)$, we have $\braket{:\! \mathcal{O}(b,b^\dag) \!:} = \int d\beta\, P(\beta) \mathcal{O}(\beta,\beta^*)$. This normal ordering will be crucial in what follows. Note that $\braket{\phi} = \braket{:\!\phi\!:}$ since $\phi \sim b^\dag b$ is already normal ordered. 

Again, for a general $P(\beta)$, Eq. \eqref{pN-q} is not the Poisson distribution. The deviations can be even more striking than in the classical case. Given Eq.~\eqref{pN-q}, one can proceed as above, and finds in similar fashion
\bea
\overline{N}&= \eta T \!\braket{\phi}\!, \\
\Delta N^2 &= \overline{N} + \eta^2 T^2 \int d\beta\, P(\beta) \left[ \phi(\beta) - \braket{\phi} \right]^2\!,
\eea
with the second result confirming Eq.~\eqref{eq:dn2-q}. As the integral is now performed over $P(\beta)$, which can take on negative values, $\Delta N^2 - \overline{N}$ can take on either sign, and exhibit sub-Poisson statistics, which are strictly forbidden for a purely classical incident state. An explicit example of this was provided in App.~\ref{app:nonclassical-example}. The result cannot be arbitrarily negative, however. As $\phi$ is proportional to the number operator,
\bea
\Delta N^2 &= \overline{N} + \eta^2 T^2 \left[ \langle :\! \phi^2 \!: \rangle - \langle :\! \phi \!:\rangle^2 \right] \\
&= \bar{N}(1-\eta T \ell^2/4 L^3) + \eta^2 T^2 \braket{\Delta \phi^2} \\
&\geq \bar{N}(1-\eta T \ell^2/4 L^3).
\eea
As expected, the size of the deviation is controlled by $\eta$.

\subsection{Amplitude variance}

Consider next the amplitude detector discussed in Sec.~\ref{sec:amplitude}. We will compute its output noise spectrum,
\be
\Delta X^2 = \overline{X^2} - \overline{X}^2
\ee
in the presence of some incoming gravitational radiation. Here the overline denotes an expectation value taken over $p(X)$. We will consider the classical and quantum calculations as well as the fixed and variable incident flux scenarios simultaneously. We take the interaction as in Eq.~\eqref{eq:V-main}, namely
\be
V = i g (a^\dag b - a b^\dag),
\ee
where for the time being $b,\, b^\dag$ can be either c-numbers or operators. To compute the variance, we will need the Krauss operators and POVM to second order in $g \delta t \ll 1$. The Krauss operator to this order is
\bea
K_{X} & = \braket{X|0} + g \delta t \braket{X | a^\dag | 0} b  \\
& + \frac{1}{2} g^2 \delta t^2 \left[ \braket{X | a^{\dag 2} | 0} b^2 - \braket{X | a a^\dag | 0} b^\dag b \right]\!.
\eea
Note again that $\ket{0}$ means the vacuum $\ket{n = 0}$, not the position eigenstate $\ket{X = 0}$. This gives the POVM elements to the same order:
\begin{align}
E_X & = K_X^\dag K_X \nonumber \\
& = \braket{0 | X} \braket{ X | 0} \nonumber \\
& + g \delta t \left[ \braket{0 | X} \braket{X | a^\dag | 0} b + \braket{0 | a | X} \braket{X | 0} b^\dag \right]  \nonumber \\
& + \frac{1}{2} g^2 \delta t^2 \left[ \braket{0 | X} \braket{X | a^{\dag 2} | 0} b^2 - \braket{0 | X} \braket{X | a a^\dag | 0} b^\dag b \right] \nonumber \\
& + \frac{1}{2} g^2 \delta t^2 \left[ \braket{0 | a^2 | X} \braket{X | 0} b^{\dag 2} - \braket{0 | a^\dag a | X} \braket{X | 0} b^\dag b \right] \nonumber \\
& + g^2 \delta t^2 \left[ \braket{ 0 | a | X} \braket{X | a^\dag | 0} b^\dag b \right]\!.
\label{eq:povm-var-app}
\end{align}
At this stage, the dependence on the detector is reduced to numerical factors involving matrix elements of creation and annihilation operators $a, a^\dag$. The dependence on the gravitational signal enters through the $b, b^\dag$. 

Now we can assume some state (either classical or quantum mechanical) for the gravitational signal, and use it to compute the average value of the detector amplitude output $\overline{X}$, by
\be
\overline{X} = \int dX \braket{E_X} X.
\ee
This follows as in the notation adopted, $p(X) = \braket{E_X}$, which is the appropriate generalization of Eq.~\eqref{eq:pofX}. The integral over $X$ reduces the inner products in Eq.~\eqref{eq:povm-var-app} to simple numbers, via manipulations of the form
\be
\int dX \braket{0 | X} \braket{X | 0} X = \braket{0 | X | 0} = 0.
\ee
One finds after straightforward algebra of this type that
\be
\overline{X} = \frac{1}{\sqrt{2}} g \delta t \braket{ b+ b^\dag} = g \delta t \braket{X_b}.
\ee
This final expectation value can be treated either classically, by integrating over a distribution $P_{\rm cl}(b)$, or quantum mechanically, by integrating in the Glauber representation $P(\beta)$. A similar calculation results in
\be
\overline{X^2} = \frac{1}{2} + \frac{1}{2} g^2 \delta t^2 \braket{ b^2 + b^{\dag 2} + 2 b^\dag b }.
\ee
This result is valid both classically and quantum mechanically; to obtain it we did not need to invoke any commutator between $b,\,b^\dag$. The first $1/2$ term comes from the commutator $[a,a^\dag]=1$, i.e., the detector's vacuum fluctuations in the quadrature variables $X,Y$.

The key difference between quantum and classical radiation variables now appears when we try to re-express this result in terms of $X_b = (b + b^\dag)/\sqrt{2}$. In the quantum case we need to use a commutator $[b,b^\dag] = 1$, and in particular one has
\bea
\overline{X^2}_{\hspace{-0.19cm}\rm cl} & = \frac{1}{2} + g^2 \delta t^2 \braket{ X_b^2 }\!, \\
\overline{X^2}_{\hspace{-0.19cm}\rm qu} & = \frac{1}{2} + g^2 \delta t^2 \left[ \braket{ X_b^2 } - \frac{1}{2} \right]\!.
\eea
Putting these results together, we obtain the variance in the as
\be
\label{eq:quadvariance-clqu}
\Delta X^2 = \frac{1}{2} + g^2 \delta t^2 
\begin{cases}  \braket{ \Delta X_b^2}\!, & {\rm classical}, \\
\big[ \!\braket{ \Delta X_b^2}\! - \frac{1}{2} \big], & {\rm quantum}.
\end{cases}
\ee
In terms of explicit integration over a classical probability or quantum Glauber representation,
\be
\Delta X^2 = \frac{1}{2} + 2 g^2 \delta t^2 
\begin{cases}  \int db\, P_{\rm cl}(b) \left[ {\rm Re}~b - \braket{ {\rm Re}~b}\right]^2 \!, & {\rm cl}. \\
\int d\beta\, P(\beta) \left[ {\rm Re}~\beta - \braket{ {\rm Re}~\beta}\right]^2\!, & {\rm qu}.
\end{cases}
\ee
In the classical case the expectation value is taken over $P_{\rm cl}(b)$, and therefore $\Delta X^2 \geq 1/2$. In the quantum case, the integral is over $P(\beta)$, which again can allow us to evade the classical bound. However, from Eq.~\eqref{eq:quadvariance-clqu} we see that even for an extremely squeezed gravitational state that achieves $\braket{\Delta X_b^2} \to 0$, we only have demonstrably quantum behavior in the range $1 > 2 \Delta X^2 \geq 1-g^2\delta t^2$. Again, although we see a distinctly quantum state is required, it is insufficient: we will also need a detector capable of registering the $g^2 \delta t^2$ effect. Notice that a large-amplitude coherent state does not make this easier.

\end{document}